\documentclass{aa}  

\usepackage{graphicx}
\usepackage{txfonts}
\usepackage{lipsum}
\usepackage{subcaption}
\usepackage{lscape}
\usepackage{placeins} 
\usepackage{ae,aecompl}
\usepackage[dvipsnames]{xcolor}
\usepackage{amsmath}
\usepackage{amssymb}
\usepackage{savesym}
\savesymbol{tablenum}
\usepackage{makecell}
\usepackage{siunitx}
\usepackage{xspace}
\usepackage{listings}
\usepackage{soul}
\usepackage{float}
\restoresymbol{SIX}{tablenum}
\usepackage{hyperref}
\usepackage{comment}
\usepackage{makecell}
\usepackage{booktabs}

\usepackage[T1]{fontenc}
\usepackage{fontawesome}
\definecolor{linkcolor}{HTML}{2AAD2E}

\usepackage{newtxtext,newtxmath}

\usepackage{multirow}
\usepackage{caption}
\usepackage{subcaption}
\usepackage{graphicx}
\usepackage{float}

\makeatletter
\newcommand{\displayappendix}{
    \renewcommand{\@seccntformat}[1]{
      \ifcsname pref@##1\endcsname \csname pref@##1\endcsname \else \csname the##1\endcsname\quad \fi
    }
    \newcommand{\pref@section}{Appendix \thesection: }
}
\makeatother

\begin{document}

   \title{The multi-wavelength vertical structure of the archetypal $\beta$~Pictoris debris disk}

    \author{Yinuo~Han\inst{1}\fnmsep\thanks{Corresponding author: yinuo@caltech.edu}
        \and Mark~C.~Wyatt\inst{2}
        \and Marija~R.~Jankovic\inst{3}
        \and Andrew~Zhang\inst{4}
        \and William~R.~F.~Dent\inst{5}
        \and \\A.~Meredith~Hughes\inst{6}
        \and Luca~Matr\`a\inst{7}
        }

   \institute{Division of Geological and Planetary Sciences, California Institute of Technology, 1200 E. California Blvd., 91125, Pasadena, CA, USA
   \and Institute of Astronomy, University of Cambridge, Madingley Road, Cambridge CB3 0HA, UK
   \and Institute of Physics Belgrade, University of Belgrade, Pregrevica 118, 11080 Belgrade, Serbia
   \and Department of Physics, University of Oxford, Oxford OX1 3RH, UK
   \and ALMA JAO, Alonso de Cordova 3107, Casilla 763 0355, Santiago, Chile
   \and Department of Astronomy, Van Vleck Observatory, Wesleyan University, 96 Foss Hill Dr., Middletown, CT, 06459, USA
   \and School of Physics, Trinity College Dublin, the University of Dublin, College Green, Dublin 2, Ireland}

   \date{Received March 2026}
 
  \abstract
   {Thermal imaging of debris disks is resolving the vertical height in an increasing number of systems, enabling the use of the vertical structure to decode the dynamical state of the planetary system. }
   {In this study, we examine the multi-wavelength structure of the archetypical edge-on debris disk of $\beta$~Pic, extensive imaging of which across mid-infrared to millimeter wavelengths makes it the prime system to study the vertical height across different grain size populations. } 
   {We non-parametrically modelled the radial profiles and constrained the vertical height at each wavelength while taking into account the vertical warping, finding the disk to be on average 1.5 times thicker vertically in the mid-infrared compared to the millimeter and the scale height to be relatively constant across radius. 
    }
   {The decreasing scale height with wavelength is in contrast to predictions from collisional damping, and could be a result of the combined effect of radiation pressure and random collisions. We also show that the disk is warped at millimeter wavelengths and find tentative evidence for clumps in ALMA images which will require follow-up observations to confirm. 
    }
   {The millimeter vertical warping is consistent with findings in scattered light and the secular perturbation interpretation due to the inner giant planets, which could also explain the relatively constant apparent scale height across radius, and potentially earlier findings of a non-Gaussian vertical profile which this study confirms.}


   \maketitle
   
   \nolinenumbers

\section{Introduction}
\label{sec:introduction}
Debris disks are extrasolar analogues of the Solar System’s Kuiper and asteroid belts. While their masses are dominated by planetesimals within the disk, mutual collisions between planetesimals feed a population of dust that is observable through its thermal emission and scattered stellar light \citep{Hughes2018}. Observations of debris disks have revealed a wide range of interesting structures, including rings and radial gaps, azimuthal asymmetries and vertical warping (e.g., \citealp{Crotts2024, Engler2025, Marino2026, Han2026b, Zawadzki2026, Lovell2026}), all of which have been used to place constraints on the outer planetary system architecture and its evolutionary history (e.g., \citealp{Marino2019, Lovell2021, Booth2023, Han2026}). 

While inferring planetary properties from disk structures is an exciting prospect, physical processes intrinsic to a debris disk also affect its evolution even in the absence of planets. For example, collisions between disk material play a central role in its evolution: bodies in a debris disk from planetesimals to micron-sized dust grains are thought to exist in a ``collisional cascade'', in which random collisions cause the largest bodies to grind down into successively smaller bodies, until dust grains at the smallest end of the size distribution are removed from the system by stellar radiation pressure \citep{Dohnanyi1969, Wyatt2008}. Since collisional kinematics may vary across grain size and spatial regions in the disk, the resulting disk structure is also expected to vary depending on the grain size population being observed. Understanding intrinsic disk dynamics is therefore paramount to making inferences with the additional complexity involving perturbing planets. 

Several forces in the absence of perturbing planets are thought to influence the orbital evolution of constituent bodies in debris disks. 
Random encounters between bodies orbiting in a central potential tend to increase their overall kinetic energy in a process known as ``viscous stirring'' (e.g., \citealp{Ida1990, Shannon2012}). 
Such a process is also accompanied by the effect of ``dynamical friction'', in which kinetic energy is preferentially transferred from larger planetesimals to smaller ones (e.g., \citealp{Ida1990, Lissauer1993}).
However, an opposing effect may be present due to ``collisional damping'' (e.g., \citealp{Brahic1977, Hornung1985, Pan2012}), in which inelastic collisions between bodies in the disk reduce their velocity dispersion, which could cause smaller grains produced from these collisions to inherit a smaller velocity dispersion compared to the larger parent grains. 
For sufficiently small dust grains, radiative forces including stellar radiation pressure become important, placing them on highly eccentric orbits upon their production, which, in combination with collisions, tend to increase their inclinations relative to their parent bodies \citep{Thebault2009}. For the smallest grains in the disk, typically micron-sized, radiation pressure exceeds half of the magnitude of stellar gravity, putting them on unbound hyperbolic orbits and ejecting them from the system. 

Observationally probing these dynamical forces acting on bodies in the disk is often difficult. However, the vertical height of debris disks may offer an important clue, as the velocity dispersion of disk material is directly linked to their inclination dispersion and thus vertical height \citep{Wyatt2002}. 
Furthermore, it is possible to constrain the inclination dispersion for different grain sizes by observing the disk at corresponding wavelengths \citep{Hughes2018}. Thermal emission from dust grains is only efficient at wavelengths similar to or smaller than the grain size, whereas the size distribution in debris disks implies that smaller grains collectively possess a significantly larger surface area than larger grains \citep{Wyatt2008}. The result of the two effects combined is that observations at a particular wavelength are broadly sensitive to grains of a comparable size to that wavelength. Multi-wavelength vertically resolved imaging could therefore allow us to observationally constrain and compare the scale height of dust populations with different grain sizes. 
Such comparisons offer observational tests to the important aforementioned forces shaping collisional evolution. For example, if collisional damping were to be a dominant effect, larger grains in the disk would be expected to exhibit a larger scale height than smaller grains \citep{Pan2012}. 

In practice, however, multi-wavelength vertical height comparisons have been difficult to make. In addition to the edge-on viewing geometry required to robustly constrain the scale height, the disk must also be vertically resolved at multiple widely separated wavelengths. With current instrumentation, the most feasible wavelengths in which to vertically resolve edge-on disks are in scattered light observed at optical/near-infrared wavelengths, due to the small point spread function (PSF) at shorter wavelengths (\citealp{Olofsson2016} and references therein), and in thermal emission at millimeter wavelengths with the Atacama Large Millimeter Array (ALMA), which offers sufficiently long baseline coverage to achieve the high angular resolution required (e.g., \citealp{Daley2019, Matra2019, Han2022, Vizgan2022, Hales2022, Marshall2023, Terrill2023, Han2025, Zawadzki2026}). 

Although scattered light observations offer high angular resolution, recovering the disk structure at these wavelengths is generally difficult due to uncertainties and observational artefacts associated with PSF subtraction, as well as the degeneracy between the scattering phase function and the intrinsic structure of the disk. In contrast, the disk structure is more readily recoverable in thermal emission, which are not subject to the effects of the scattering phase function. 

Studies that model the vertical structure in debris disks commonly assume the scale height to be proportional to radius, resulting in a constant aspect ratio, $h$, which can be fitted as a free parameter (e.g., \citealp{Daley2019}). It is also possible to drop this assumption and attempt to recover any radial variations in the scale height. This could be achieved through a non-parametric approach (e.g., \citealt{Han2022}), which avoids assuming a parametrisation of the scale height as a function of radius, or a parametric approach (e.g., \citealp{Matra2019, Terrill2023}). Observations with high resolution and sensitivity are required for such modelling to yield robust results. 

Focusing on thermal emission observations, a systematic comparison of debris disk scale heights between wavelengths has only been carried out for the edge-on debris disk of AU~Mic \citep{Vizgan2022}, in which vertically resolved imaging at two ALMA wavelengths (Band 6 at 1.3~mm and Band 8 at 0.45~mm) were studied. Although the wavelength separation was modest---both being within the sub-millimeter regime---the study found the height to be smaller at the shorter wavelength band by a factor of 1.35$\pm$0.09, which would be a remarkably steep dependence.  

A key to testing any grain size dependence of the scale height lies in sensitive observations at more widely separated wavelengths. This is generally difficult to achieve because vertically resolving debris disks at mid-infrared wavelengths---the shorter wavelength end of thermal emission observations---is severely limited by resolution and sensitivity constraints. However, the most notable exception is the edge-on debris disk of the nearby A-star $\beta$~Pic for which, due to its proximity (19.63~pc, \citealt{Gaia2018}), luminosity and young age (21~Myr, \citealt{Binks2014}), there exist multiple epochs of well-resolved mid-infrared images at multiple wavelengths between 8 and 25 microns, as well as high-resolution and sensitivity ALMA observations across multiple bands. $\beta$~Pic therefore offers a rare opportunity to study the vertical height of a debris disk across very widely separated wavelengths. 

To characterise the vertical height of the $\beta$~Pic debris disk, various substructures found in the disk must be considered. Most notably, the disk exhibits a vertical warp observed in scattered light \citep{Heap2000, Golimowski2006, Apai2015}, where the maximum projected vertical displacement from the disk midplane is reached at 70--80 au from the star, corresponding to a projected tilt of 4--5$^{\circ}$ from north to east (i.e., counterclockwise) from the disk midplane \citep{Apai2015}. The warp is thought to be caused by gravitational perturbations from inclined planets \citep{Mouillet1997}, which subsequently detected planets, $\beta$~Pic~b (12$\pm$3\,M$_{\text{Jup}}$ at 9.9\,au, \citealp{Lagrange2009, Lacour2021}) and c (9$\pm$1\,M$_{\text{Jup}}$ at 2.7\,au, \citealp{Nowak2020, Lacour2021}) are able to achieve \citep{Nesvold2015, Smallwood2023}. Another prominent structure is a clump of dust observed in the mid-infrared to extend from 40 to 80\,au from the star along the SW arm \citep{Telesco2005, Li2012, Han2023}, and a co-located clump of gas observed by ALMA that has been best resolved in molecular CO \citep{Dent2014, Matra2017} and tilted from the midplane of the dust disk by an amount consistent with the aforementioned warp. Similar asymmetries have also been observed in atomic carbon by ALMA \citep{Cataldi2018}. The nature of this prominent dust and gas clump remains unclear, but explanations including a giant collision (e.g., \citealp{Telesco2005, Han2023, Rebollido2024}), tidal disruption \citep{Cataldi2018}, resonant planetesimals \citep{Wyatt2003, Dent2014}, secular perturbation \citep{Nesvold2015}, collisional avalanche \citep{Grigorieva2007} and gas vortex \citep{Skaf2023} have been suggested. 

While these substructures are interesting topics of investigation, they also mean that one must take into account their presence when modelling and interpreting the vertical height. 
Nonetheless, $\beta$~Pic currently still offers the best opportunity to systematically study the multi-wavelength scale height in any debris disk to piece together robust observational evidence of scale height variations with grain size and constrain models on the intrinsic collisional evolution in debris disks. 

This study therefore aims to combine the available high-resolution imagery of $\beta$~Pic at mid-infrared and millimeter wavelengths to investigate the multi-wavelength vertical structure of $\beta$~Pic and shed light on whether dynamical processes differentially affect different grains sizes. Section~\ref{sec:obs} describes the observations used in this study. We fit the radial structure and construct the spatially resolved SED of different radial regions of the disk in Section~\ref{sec:radial}, before building on these radial profiles to measure the vertical height in Section~\ref{sec:vertical}. Section~\ref{sec:discussion} discusses implications on collisional processes affecting dust grain evolution in debris disks more generally as well as on the dynamical scenario shaping $\beta$~Pic. Our findings are summarised in Section~\ref{sec:conclusions}.

\section{Observations and data reduction}
\label{sec:obs}

\begin{figure}
    \centering
    \includegraphics[width=8.5cm]{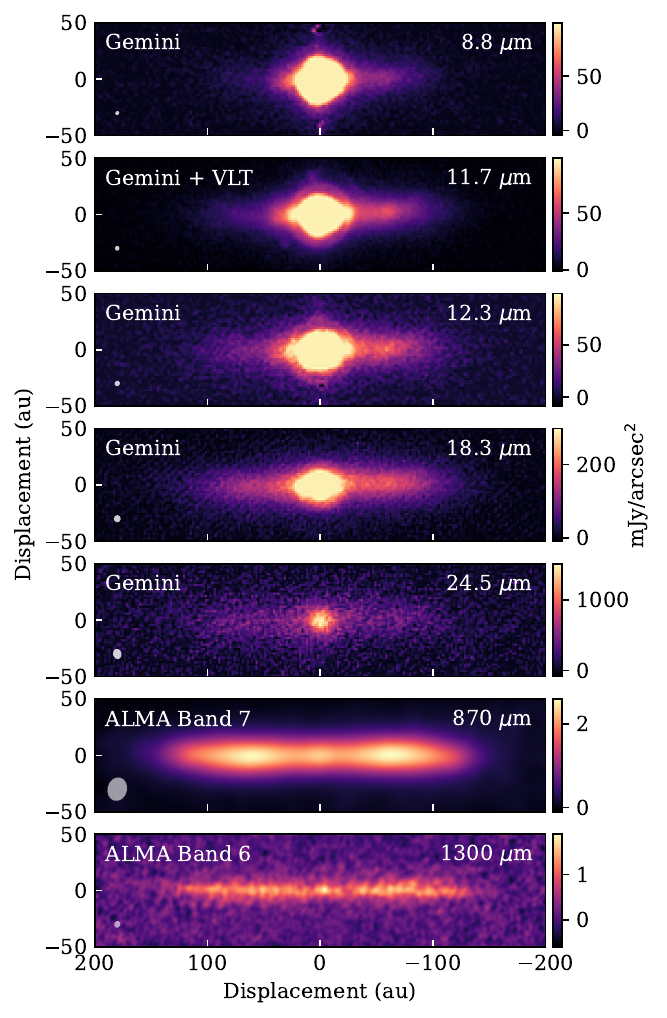}
    \caption{Imagery of $\beta$~Pic used in this study. The mid-infrared images were taken by Gemini/T-ReCS \citep{Telesco2005, Li2012} and VLT/VISIR \citep{Han2023}. The millimeter dust continuum images were observed with ALMA Band 7 \citep{Hull2022} and Band 6 \citep{Matra2019}. The mid-infrared images are displayed on a linear scale capped at 0.12, 0.12, 0.12, 0.37 and 1.9 Jy/arcsec$^2$ respectively from the shortest to longest wavelength. The ellipses indicate the half-maximum contour of the PSF (mid-infrared) or synthetic beam (ALMA) when modelled as a 2D Gaussian. }
    \label{fig:obs}
\end{figure}

This study uses available images of $\beta$~Pic at 5 mid-infrared and 2 millimeter wavelengths which are summarised in Table~\ref{table:obs} and displayed in Fig.~\ref{fig:obs}. Details of the observations and data reduction are described in this section. 

\subsection{Mid-infrared imaging}
$\beta$~Pic has been imaged in the mid-infrared at 8.8, 11.7, 12.3, 18.3 and 24.5~$\mu$m by the T-ReCS instrument on the Gemini South Telescope at Cerro Pach\'on Observatory, and at 11.5~$\mu$m by the VISIR instrument on the Very Large Telescope at Paranal Observatory. There also exist coronagraphic observations as part of the NEAR experiment on VISIR \citep{Skaf2023} and by JWST/MIRI \citep{Rebollido2024}, however uncertainties associated with characterising the spatially-variant PSF and inner disk regions make it difficult to robustly model and constrain the disk's scale height in a consistent way for comparison with full aperture imagery. Given the abundance of full aperture images available in the mid-infrared, we do not include the coronagraphic datasets in this study. 

As these observations were made with different instruments and previous analyses focused on one particular instrument \citep{Telesco2005, Li2012} or wavelength \citep{Han2023}, we re-reduced all observations to ensure consistency across the mid-infrared dataset for subsequent multi-wavelength analyses.

\begin{table*}
\caption{Summary of observations used in this study. }
\label{table:obs}
\begin{tabular}{llllllll}
\hline \hline
Regime & Instrument    & Filter/band                       & Wavelength                                            & Observing date          & Resolution   & PSF ref. & Ref. \\
                  &               &                                   & ($\lambda$ / $\Delta \lambda$) [$\mu$m] &                         & {[}arcsec{]} &                    \\ \hline
Mid-infrared      & Gemini/T-ReCS & Si2-8.8um                         & 8.7 / 0.8                                             & 2 Dec 2003              & 0.31         & HD~50310     & (1) \\
                  &               & Si5-11.7um                        & 11.7 / 1.1                                            & 30 Dec 2003             & 0.38         & HD~50310  & (1)    \\
                  &               &                                   &                                                       & 16 Dec 2010             & 0.41         & HD~39523  & (2)    \\
                  &               & Si6-12.3um                        & 12.3 / 1.2                                            & 2 Dec 2003              & 0.41         & HD~50310  & (1)    \\
                  &               & Qa-18.3um & 18.3 / 1.5                                            & 2, 3, 29, 30 Dec 2003   & 0.54         & HD~50310   & (1)   \\
                  &               & Qb-24.5um & 24.6 / 2.0                                            & 3, 29 Dec 2003          & 0.72         & HD~50310    & (1)  \\
                  & VLT/VISIR     & B11.7                             & 11.5 / 0.9                                            & 1, 15 Sep 2015          & 0.38         & HD~50310    & (3)  \\ \hline
Millimeter        & ALMA          & Band 7                            & 800--1100                                             & 18, 19 Dec 2019         & 0.97         &           & (4)         \\
                  &               & Band 6                            & 1100--1400                                            & Oct, Dec 2013, Aug 2015 & 0.29         &           & (5), (6)         \\ \hline
\end{tabular}
\tablefoot{References are (1) \citealp{Telesco2005}, (2) \citealp{Li2012}, (3) \citealp{Han2023}, (4) \citealp{Hull2022}, (5) \citealp{Matra2017} and (6) \citealp{Matra2019}. }
\end{table*}

\subsubsection{Mid-infrared sky background subtraction}
All observations were carried out with a chop-nod strategy as is standard for ground-based mid-infrared imaging, in which the telescope pointing and secondary mirror periodically oscillate between ``on-source'' and ``off-source'' positions, enabling the removal of the thermal background emission from the sky and the telescope during post-processing. A detailed description of the chop-nod strategy and reduction process to remove the sky and telescope background emission is available in Section~1.3 of \citet{Han2023} for both Gemini/T-ReCS and VLT/VISIR data, and the same standard reduction procedure is applied in this study. 

For the 8.8, 11.7 and 12.3~$\mu$m observations, we re-centred each exposure frame before stacking them to correct for any inter-frame pointing offsets during the chop-nod cycle. Centring was performed by fitting a 2D Gaussian model to each frame and using the peak of the fitted Gaussian to estimate the location of the star. Such a correction could not be robustly performed for the 18.3 and 24.5~$\mu$m images due to the low S/N in individual exposure frames and the significant flux contribution from the extended disk, and we stacked all exposure frames directly without correcting for potential pointing offsets between frames. 

The Gemini/T-ReCS images exhibited a ``DC offset'' in which particular rows or columns of pixels have a constant background offset particularly when passing through the bright PSF core. We corrected for these detector read-out artefacts by subtracting off the DC offset in each row and column, which is determined using the median of the row or column of pixels after masking out emission from the source.

\subsubsection{Flux calibration}
The mid-infrared observations at each filter and epoch were accompanied by PSF reference star observations of a mid-infrared standard star as listed in Table~\ref{table:obs}. We calibrated all images into physical flux density units (i.e., Jy/arcsec$^2$) using the in-band flux densities of corresponding mid-infrared standard stars maintained by Gemini and Paranal Observatories. 

\subsubsection{Combining the 11.7\,$\mu$m observations}
$\beta$~Pic has been imaged at approximately 11.7\,$\mu$m over three epochs by Gemini \citep{Telesco2005, Li2012} and the VLT \citep{Han2023} combined.
Since the primary aim of this paper is to constrain the vertical height, subsequent analyses would benefit from the highest S/N map possible. It is known that there is a dust clump on the SW arm of the disk \citep{Telesco2005} which could possibly be moving, however its motion, if any, is likely small and within a few au in projection over the range of observing dates \citep{Han2023}. Such a small displacement, if any, is not expected to affect our analysis of the large-scale radial and vertical structure of the disk, and we will focus on the northeast (NE) arm of the mid-infrared disk without this dust clump to further mitigate its effect. 

We therefore combined the two epochs of T-ReCS imaging with the Si5-11.7~$\mu$m filter and one epoch of VISIR imaging with the B11.7 filter into a single image by re-sampling the VISIR image to the T-ReCS plate scale (down-sampling from 45~mas/pixel to 90~mas/pixel with a two-dimensional cubic spline interpolation) and taking the mean across all epochs weighted by the inverse of their respective mean-squared noise. The same re-sampling was applied to all corresponding normalised PSF observations and combined with the same weights as those for the $\beta$~Pic observations to obtain the effective PSF for the combined image. We refer to this combined image as the ``11.7~$\mu$m image'' throughout the rest of this paper.

\subsubsection{Correcting additional detector readout artefacts}
The Gemini/T-ReCS detectors are prone to artefacts when imaging bright sources in addition to DC offsets. This particularly affects the PSF reference star observations at 8.8, 11.7 and 12.3~$\mu$m, in which pixels in a given column that passes through the bright PSF core are offset by a negative or positive value on either side of the bright source respectively. Pattern noise correction is common in mid-infrared data reduction\footnote{See mid-infrared reduction package from Gemini Observatory \href{https://www.gemini.edu/sciops/data/IRAFdoc/midirinfo.html}{www.gemini.edu/sciops/data/IRAFdoc/midirinfo.html}} and has previously been applied to mid-infrared observations of $\beta$~Pic \citep{Li2012}. To better characterise the true PSF in order to more accurately disentangle disk flux in the images of $\beta$~Pic from the PSF of the star during subsequent modelling, we corrected for the positive and negative ``streaks'' on either side of the PSF reference star. This was done by shifting the baseline value in each region affected by the readout artefact that is further than 1.8$^{\prime\prime}$ from the star such that the mean becomes 0. In practice, only the central few columns of pixels need to be corrected in each image upon inspection.

\subsection{Millimeter imaging}
\subsubsection{Band 7}
\label{sec:obsband7}
Linear polarisation observations of $\beta$~Pic have been taken with ALMA at Band 7 (870~$\mu$m) at 0.97$^{\prime\prime}$ (19\,au) resolution \citep{Hull2022}. Although Band 7 observations with half this beam size also exist \citep{Dent2014}, those observations were obtained with more than twice the root-mean-squared (rms) noise per beam. We base subsequent Band 7 analyses on the \citet{Hull2022} dataset given its higher sensitivity. Due to the large angular extent of the disk, the observations were performed with a mosaic consisting of three pointings centred on the star and 3.5$^{\prime\prime}$ along either arm of the edge-on disk respectively. Further details of the observations and data reduction are described in Section~2 in \citet{Hull2022}. Using the calibrated visibilities from \citep{Hull2022}, we obtained primary beam-corrected dust continuum images with the \texttt{tclean} task in the \texttt{CASA} program \citep{casa2022} with auto-masking. We produced two Briggs-weighted versions of the image with a robust value of 2.0, one without UV-tapering and the other with UV-tapering (by $\mathrm{bmaj} = 0.8^{\prime\prime}$, $\mathrm{bmin} = 0.5^{\prime\prime}$ and $\mathrm{bpa} = 15.7^{\circ}$) to achieve a circular beam with a FWHM of 1.2$^{\prime\prime}$. The version shown in Fig.~\ref{fig:obs} is imaged without a UV-taper.

\subsubsection{Band 6}
$\beta$~Pic has also been observed in the dust continuum at Band 6 (1300~$\mu$m) at 0.29$^{\prime\prime}$ (6\,au) resolution \citep{Matra2019}. The observations were carried out with both an extended configuration under a single pointing and a compact configuration with two pointings centred at 5$^{\prime\prime}$ along either arm of the disk respectively. Details of the observations are described in Section~2.1 in \citet{Matra2017} and Section~2 in \citet{Matra2019}. We used the calibrated visibilities of this dataset made available via the REASONS sample \citep{Matra2025} and produced the same two versions of primary beam-corrected images as described in Section~\ref{sec:obsband7} (with the 1.2$^{\prime\prime}$-beam version UV-tapered by $\mathrm{bmaj} = 1.1^{\prime\prime}$, $\mathrm{bmin} = 0.85^{\prime\prime}$ and $\mathrm{bpa} = 7.6^{\circ}$). To compare between the emission at Band 6 (higher resolution) and Band 7 (lower resolution), we also produced an additional Band 6 image with a UV-taper (by $\mathrm{bmaj} = 0.77^{\prime\prime}$, $\mathrm{bmin} = 0.65^{\prime\prime}$ and $\mathrm{bpa} = 25.4^{\circ}$) that achieves the same beam size and orientation as the Band 7 image without a UV-taper. 
The version shown in Fig.~\ref{fig:obs} is imaged with a robust value of 2.0 without a UV-taper.

\subsection{Position angle of the disk}
We measured the position angle of the disk at each wavelength by finding the orientation of a narrow rectangular slit passing through the star which maximises the flux that falls within the slit. 

The slit centred on the star was chosen to measure 400\,au along its long edge, beyond which no flux is detected in any of these images even along the major axis of the disk. For the width of the short edge of the slit, there is no particular reason to chose one value over another, so we repeated the procedure over a range of slit widths, measuring the position angle of the disk for each slit width. We used the mean and standard deviation among all measurements to estimate the position angle and its uncertainties. 

For the mid-infrared images, the width of the slit was varied linearly between 3.5\,au and 67\,au at increments of 3.5\,au (2 T-ReCS pixels). To mitigate any bias on the position angle due to asymmetries in the PSF at the bright core, the emission within 26\,au from the star was masked out when optimising for the flux density that falls within the slit. For the ALMA Band 7 image, the width of the slit was varied between 2 and 40\,au at increments of 2\,au, and for the Band 6 image between 2 and 20\,au with the same increment. 

The measured position angles are shown in Table~\ref{table:fluxangle}. The measurements across the mid-infrared appear to be consistent (with a mean of $31.6\pm0.9^{\circ}$), confirming the $32.0\pm0.8^{\circ}$ position angle previously determined from the multiple epochs of observations at 11.7~$\mu$m \citep{Han2023}. However, the ALMA position angles appear to be smaller that in the mid-infrared, but consistent with previous parametric modelling in ALMA Band 6 ($29.7\pm0.1^{\circ}$, \citealp{Matra2019}). \citet{Han2023} noted that the mid-infrared position angle is also slightly larger than the scattered light measurement of $29.1\pm0.1^{\circ}$ from \citet{Apai2015}. The ALMA position angle therefore appears to be more similar to that of the scattered light than the mid-infrared. 

Assuming the disk midplane to be edge-on, these differences appear to be consistent with the effect of vertical warping, which causes inner regions of the disk to appear to be tilted counterclockwise from the midplane, increasing the effective position angle of the inner regions in projection. The position angles in both the scattered light \citep{Apai2015}, which was measured based on the brightest point at 200\,au separation from the star, and ALMA, in which the disk emission extends to almost 200\,au, could be reflective of the position angle of the overall midplane of the broader disk beyond the warp at $\sim$50--100\,au. In contrast, the mid-infrared emission probes closer-in emission and could have reflected more closely the position angle of the warped inner regions of the disk. In fact, the position angle measured at 24.5~$\mu$m, which probes further out emission than the shorter wavelengths, already shows a tentative decrease to a value closer to that of ALMA wavelengths, although the image at this wavelength has a relatively low S/N. 

The thermal background-subtracted, flux-calibrated and rotated images from the above procedures are displayed in Fig.~\ref{fig:obs}. This figure assumes a major axis position angle of 29.7$^{\circ}$ across all wavelengths for ease of cross-wavelength comparison. In the rest of this study, we emphasize structures within each wavelength relative to its midplane by aligning the disk's major axis horizontally at the respective wavelength (assuming a position angle of 32$^{\circ}$ in the mid-infrared and 29.7$^{\circ}$ in the millimeter) unless otherwise stated. For scale height fitting, a warped model will be subsequently introduced and assumed. 

\subsection{Aperture photometry}
We performed aperture photometry to measure the total flux density of the system at each wavelength. For the 11.7~$\mu$m image, VISIR's chop-nod cycle resulted in sub-images arranged in such a way that a rhombus-shaped aperture was well-suited to encircle the flux from the primary image while avoiding surrounding sub-images on the detector array. We chose the vertices of the rhombus to be the two points 177\,au along the disk's major axis on either side of the star and the two points 71\,au from the star along the minor axis. For all other mid-infrared images, we used a rectangular aperture with side lengths of 353 and 124\,au centred on the star with the long edge oriented along the disk's major axis. The stellar flux is included in the total flux density measurements and does not appear saturated as suggested by the shape of its 1D profile. For the two ALMA images, we used a similarly positioned rectangular aperture but with side lengths of 400 and 100\,au. We estimated the uncertainties of the total flux densities in the mid-infrared based on the root-mean-squared noise in a background region with no disk emission and summed the rms noise per pixel in quadrature within the aperture assuming the pixels on the detector are independent. 
The measured total flux densities are displayed in Table~\ref{table:fluxangle}.

\begin{table}
    \centering
    \caption{Measured position angles and calibrated total in-band flux densities for observations with each filter/band. }
    \label{table:fluxangle}
    \begin{tabular}{lll}
    \hline\hline
    Wavelength ($\mu$m) & Position angle ($^{\circ}$) & Total flux density (Jy) \\ \hline
    8.8   & $31.6\pm1.6$      & $3.070\pm0.003$      \\
    11.7  & $32.4\pm0.5$      & $2.7740\pm0.0013$    \\
    12.3  & $31.6\pm0.8$      & $2.725\pm0.005$      \\
    18.3  & $32.0\pm0.9$      & $5.143\pm0.013$      \\
    24.5  & $30.6\pm0.6$      & $13.72\pm0.15$       \\
    870   & $29.55\pm0.07$    & $0.053\pm0.003$ \\
    1300  & $29.3\pm0.2$      & $0.020\pm0.001$ \\
    \end{tabular}
    \tablefoot{Both the position angle and flux density may be subject to additional systematic uncertainties dependant on the instrument and detectors that are not reflected in the uncertainties estimated here. The ALMA flux densities assume a 5\% absolute calibration uncertainty. }
\end{table}

\section{Radial and azimuthal structure}
\label{sec:radial}

It is necessary to understand the disk's radial structure before further investigating the vertical structure. The primary aim of this section is to derive the radial profiles of the disk across all wavelengths. We begin by investigating the presence of any substructures before applying a non-parametric method to recover the radial profiles from the images independent of knowledge of the vertical height. We also leverage the wide wavelength coverage of the derived radial profiles to study the spatially-resolved spectral energy distribution (SED) at different radial regions in the disk. 

\subsection{Azimuthal asymmetries}
\label{sec:asymmetries}

Before modelling the radial profile of $\beta$~Pic, it is important to take note of the multiple substructures in the disk. Several radial and azimuthal substructures have already been identified in $\beta$~Pic (we defer the discussion on vertical substructures to Section~\ref{sec:vertical}). A prominent dust clump is known in the SW arm of the disk visible at all 5 mid-infrared wavelengths \citep{Telesco2005}, extending from 40 to 80\,au from the star. The SW arm also appears to be brighter and more extended in the mid-infrared, as is increasingly obvious towards the shorter wavelengths due to the dust clump. 

The clump has also been detected in CO observations with ALMA \citep{Dent2014, Matra2017}, which showed the CO disk to be tilted from the dust disk's midplane seen in ALMA continuum observations by less than $<5^{\circ}$ in the same direction as the scattered light warp. However, the presence of this clump has not been identified in the dust continuum at millimeter wavelengths. The most recent ALMA studies in Band 7 \citep{Hull2022} and Band 6 \citep{Matra2019} did not report any significant brightness asymmetries between the two arms, which is in contrast to suggestions from earlier ALMA Band 7 data \citep{Dent2014}, which showed the SW arm to be 15\% brighter than the NE arm between 30 and 80\,au from the star. 

In addition, mid-infrared coronagraphic observations have suggested the presence of a secondary dust clump on the NE side of the disk extending from 30 to 40\,au from the star, although significantly less prominent than the SW clump \citep{Skaf2023}. This secondary clump has not been identified at millimeter wavelengths. 

\begin{figure}
    \centering
    \includegraphics[width=8.5cm]{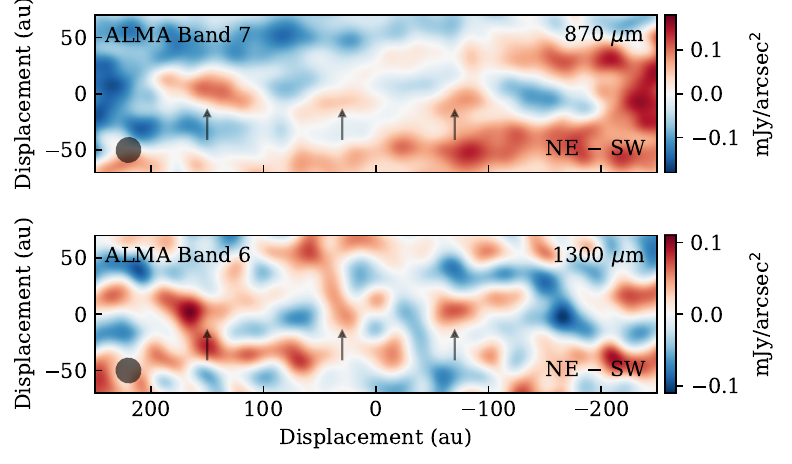}
    \caption{Rotationally subtracted ALMA images of $\beta$~Pic obtained by subtracting each image rotated by 180$^\circ$ about the star from the original image to emphasize any asymmetric features in the disk. Both images were UV-tapered to achieve an effective beam FWHM of 23\,au for both images, as indicated by the ellipses. 
    Arrows are drawn at 30, -70 and 150\,au from the star along the major axis. }
    \label{fig:image180_alma}
\end{figure}

To further investigate the presence of radial and azimuthal asymmetries in the disk, we subtracted off disk emission at each location by the emission at the opposite location about the star to obtain a ``rotationally subtracted'' image at each wavelength, as displayed in Fig.~\ref{fig:image180_alma} and Fig.~\ref{fig:image180_mir}. The images were centred by miniminising the rotationally subtracted residuals.

\begin{figure}
    \centering
    \includegraphics[width=8.5cm]{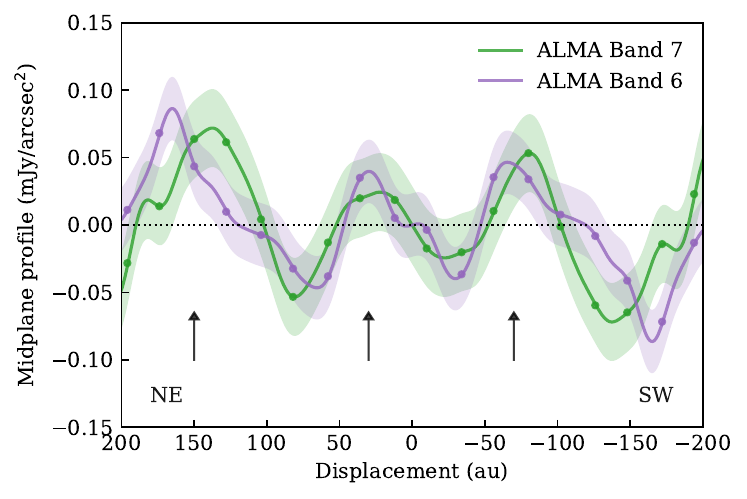}
    \caption{Mean projected surface brightness profile of the smoothed and rotationally subtracted ALMA images displayed in Fig.~\ref{fig:image180_alma} within 15\,au from the disk's major axis as a function of projected separation. Sample points separated by the effective (smoothed) beam FWHM are plotted to indicate the spatial correlation scale. The uncertainties plotted correspond to those for individual binned regions indicated by the sample points and are estimated based on the rms noise per beam of each image. The arrows are at the same location as those shown in Fig.~\ref{fig:image180_alma}. }
    \label{fig:image180_alma2}
\end{figure}

\begin{figure}
    \centering
    \includegraphics[width=8.5cm]{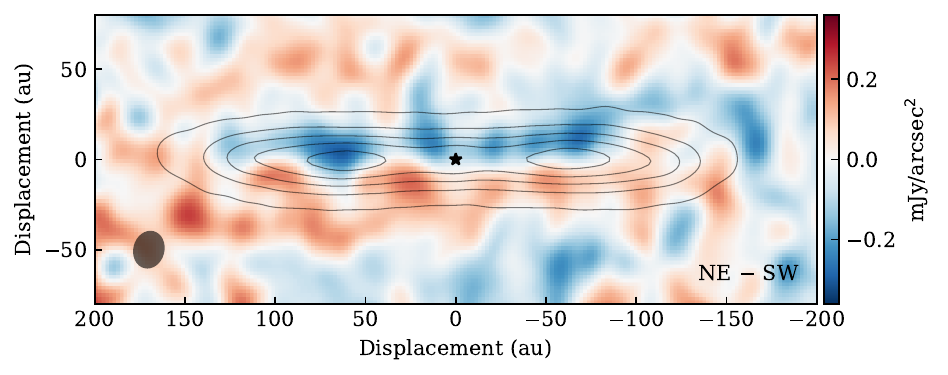}
    \caption{Difference image obtained by subtracting the Band 7 image from the Band 6 image rescaled to the Band 7 flux. Both images have been imaged to achieve the same resolution (with a robust value of 2.0 and the Band 6 image UV-tapered to match Band 7). Contours of the Band 7 image are drawn at 0.3, 0.8, 1.3, 1.8 and 2.3 mJy/arcsec$^2$. The beam is shown in the lower left corner.}
    \label{fig:alma_diff}
\end{figure}

\begin{figure*}
    \centering
    \includegraphics[width=18cm]{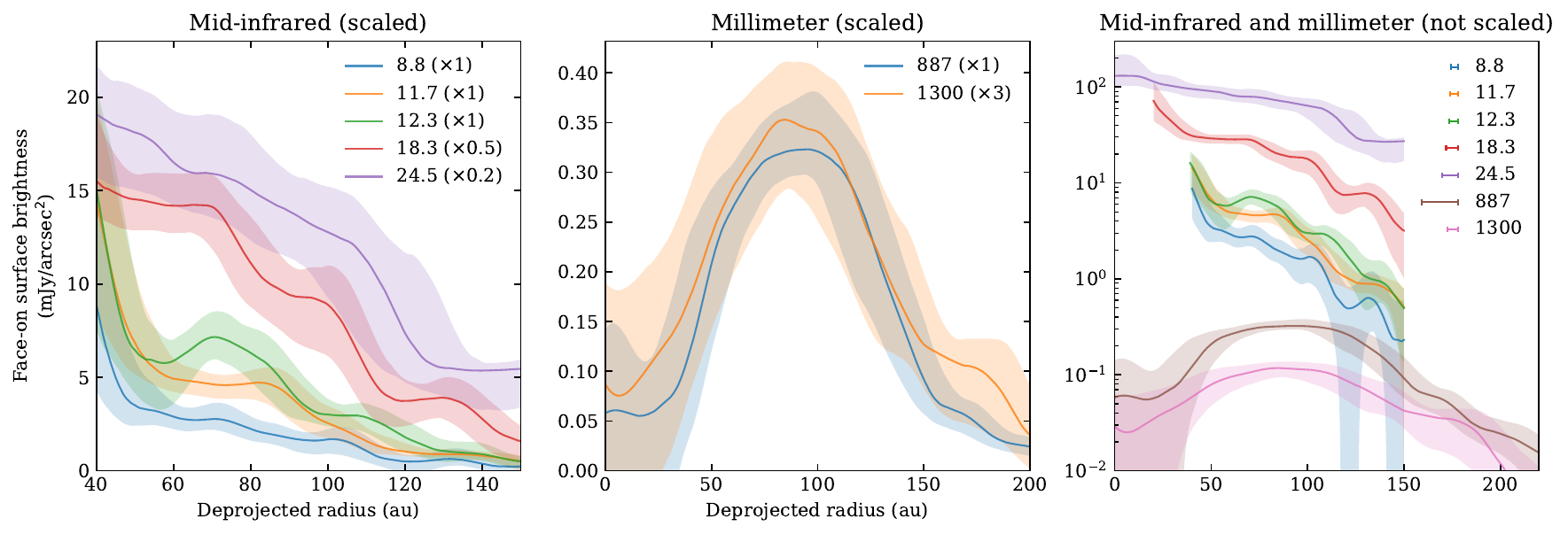}
    \caption{De-projected and deconvolved radial surface brightness profiles of $\beta$~Pic fitted non-parametrically with \texttt{rave}. The left and centre panels display the radial profiles in the mid-infrared and millimeter respectively, with each profile vertically scaled by the factor shown in parentheses in the legend following the wavelength in $\mu$m. The right panel displays the same radial profiles but on a logarithmic scale without scaling; the length of each line segment in the legend indicates the FWHM of the PSF/beam}. The shaded regions indicate the 1$\sigma$ range of possible models, within which a profile can be drawn while still reasonably reproducing the data. The corresponding fitted central component (star + potential unresolved inner dust) fluxes are shown in Table~\ref{tab:starflux}, which are not displayed in the radial profiles shown. From the shortest to longest wavelength, the number of annuli used in the fitting are 10, 10, 10, 10, 5, 10 and 7 respectively, which were chosen based on the S/N and resolution of each image. 
    \label{fig:radial_profiles}
\end{figure*}

At first glance, the blobby features in the self-subtracted UV-tapered ALMA images in Fig.~\ref{fig:image180_alma} appear to be consistent with noise. 
However, at three locations along the midplane with known mid-infrared or scattered light features, there appear to be corresponding positive residual blobs along the disk's midplane in the Band 7 self-subtracted image (and corresponding negative blobs by definition of self-subtraction). Interestingly, although the Band 6 image is noisier, positive noise-like blobs are seen at near-identical locations, although each of them is certainly insignificant in this dataset alone. While it is tempting to conclude that the residual image is featureless, we attempted to compare these blobs with known substructures at these locations. 

To do so, we plotted an emission profile along the disk's midplane by averaging the emission within 15\,au above and below the midplane, as shown in Fig.~\ref{fig:image180_alma2}. While no feature alone is confidently detected in either band, there is again a close match in these non-significant features between both ALMA bands and with known mid-infrared features. 

Firstly, we observe that the NE side of the disk appears to be slightly brighter near the outer edge of the ALMA emission, approximately between 100 and 200\,au. 
This disk region is not probed within the sensitivity of ground-based mid-infrared instruments, but is consistent with the longer NE arms observed in scattered light \citep{Larwood2001}. There may also be evidence of this feature from recent observations using the MIRI Imager on JWST, in which a tilted ``secondary disk'' is found to extend out to this distance on the NE arm (see Fig.~1 in \citealp{Rebollido2024}). The MIRI images also detected a filamentary structure named the ``cat's tail'' in the SW arm, but it extends away from the major axis at a large angle such that it would not have contributed significant flux in the line profile even if it were to be detected in the ALMA observations used here. Note that the SW arm being more extended in ground-based mid-infrared observations is contrary to this finding, possibly because the ground-based observations are sensitive enough to detect the SW dust clump, but not sensitive enough to detect the further extended structures including the tilted secondary disk to the NE. 

Secondly, co-located with the mid-infrared clump at 40 to 80\,au and the CO clump at 30 to 90\,au, enhanced Band 7 emission appears to exist between 50 to 100\,au from the star on the SW side of the side. This feature could represent a tentative detection of the dust clump in the millimeter, however the significance is low.

Thirdly, at the location of a secondary dust clump at $\sim$35\,au based on mid-infrared coronagraphic imaging with adaptive optics \citep{Skaf2023}, we see an additional region of subtle enhanced emission peaking at 30\,au from the star on the NE arm in Band 7. It is possible that these are the same potential features, however the significance is even lower than the other potential features discussed above. 

Given the different observing bands, observing dates and independent data reduction, these low-amplitude residuals could potentially be real features. 
However, it is important to caveat that the Band 7 disk emission is surrounded by subtle residual background features, which we were not able to remove by cleaning more deeply, making the residual blobs within the disk more difficult to quantify reliably. As noted in \citet{Hull2022}, positive and negative ripples in the image (larger that the beam and comparable to the disk size) suggest the presence of large-scale emission that are not well-characterised by these observations. 

To search for any differences in the spatial distribution of emission between the two ALMA bands, we subtracted the image in Band 7 from Band 6, as shown in Fig.~\ref{fig:alma_diff}. Note that the two images have been UV-tapered to achieve the same beam size and scaled to the same flux before subtraction for comparability. The difference image in Fig.~\ref{fig:alma_diff} does not reveal significant structural differences between the two bands. From the image, it is possible that the Band 6 NE peak is under-luminous relative to Band 7, which could be the case if the SW peak exhibits a shallower sub-millimeter spectral index in the region of the known SW dust clump compared to the NE side. However, given the low significance and lack of prior supporting observational evidence, we are unable to draw such a conclusion. 

Turning our attention to the mid-infrared, the rotationally subtracted mid-infrared images and corresponding line profiles are shown in Appendix~\ref{appendix}. The most prominent feature is the SW clump already known from prior studies. There appears to be marginal evidence for the secondary clump, though not as clearly as in the coronagraphic images \citep{Skaf2023} which aimed to achieve superior sensitivity and PSF stability by experimenting with the addition of an adaptive optics system and coronagraph to the standard VLT/VISIR camera.

We conclude that, apart from the SW clump in mid-infrared, we cannot reliably confirm any other asymmetries. The asymmetric ALMA features described here are not significant from each observation alone, and the possibility of their existence is primarily supported by their consistent presence in the two different sets of ALMA observations and close correspondence to known features (e.g., primary and secondary dust clump and more extended NE disk) in scattered light images and at mid-infrared wavelengths, which set a strong prior. We suggest future high-sensitivity follow-up observations to verify these structures in the millimeter. 

Apart from the SW clump, these asymmetries, even if real, are too low in amplitude to affect our aim of recovering the underlying radial structure of the disk. We therefore proceed to fit models without accounting for these features, assuming instead that the NE arm of the disk represents an azimuthally symmetric underlying distribution at mid-infrared wavelengths and that both arms of the disk are symmetric about the star at ALMA wavelengths. Any small-amplitude residual structures in subsequent model-fitting (Section~\ref{sec:almamodel}) should take into account the fact that they may correspond to these subtle asymmetries intrinsic to the observations, which could be checked by comparison with Figs.~\ref{fig:image180_alma} and \ref{fig:image180_mir}. 

\begin{figure*}
    \centering
    \includegraphics[width=17cm]{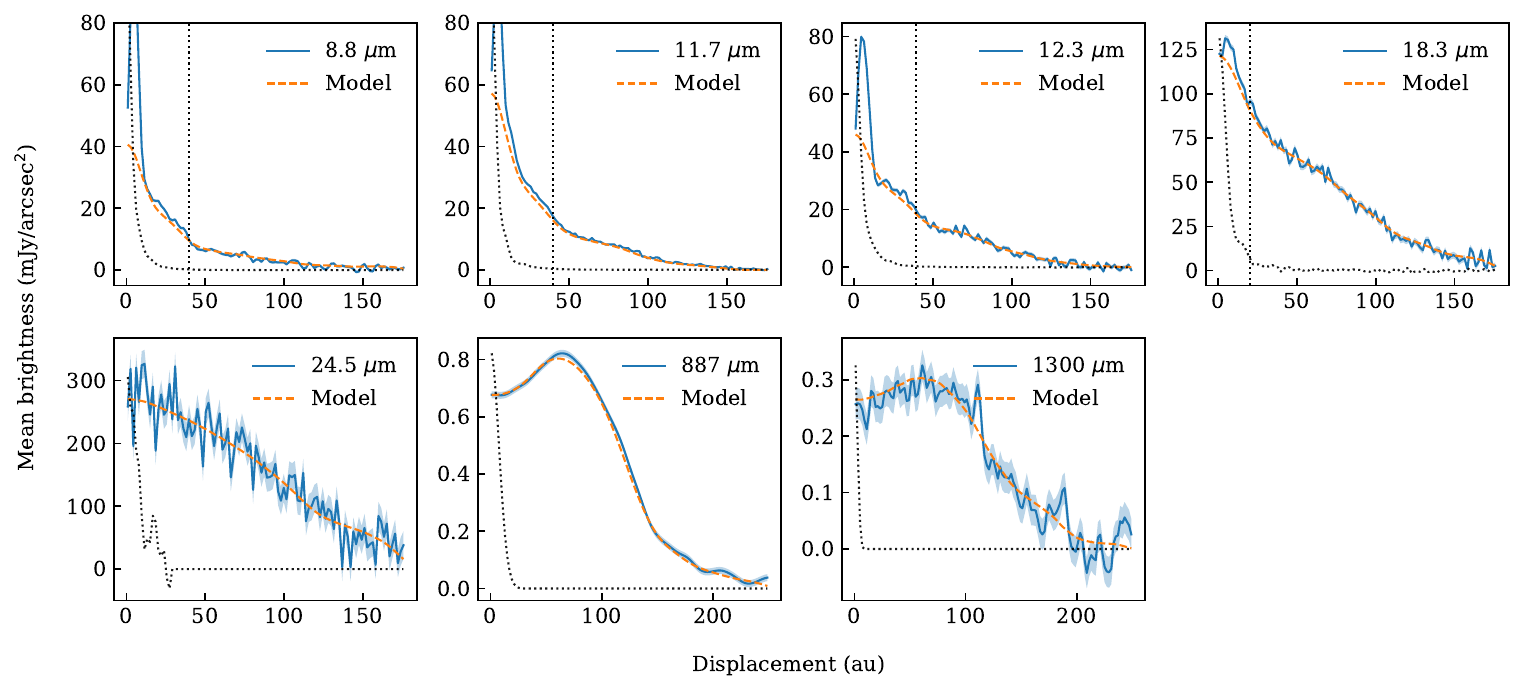}
    \caption{One-dimensional flux of each image and fitted model obtained by summing the disk emission onto the disk's major axis. This represents the quantity that \texttt{rave} is fitting to. The dotted profiles indicate the PSF/beam. Due to limitations in the PSF stability in the mid-infrared, the bright stellar core is difficult to subtract perfectly, and dotted vertical lines are drawn at the shorter wavelengths to indicate the region within which the derived radial profile is unreliable. These boundaries are determined to be 40, 40, 39 and 20\,au from the star at 8.8, 11.7, 12.3 and 18.3~$\mu$m respectively. Note that the mid-infrared uncertainties were estimated using the background noise level and assuming pixels are independent. The true noise level could be higher due to correlated noise.}
    \label{fig:1D_residuals}
\end{figure*}

\begin{figure}
    \centering
    \includegraphics[width=8.5cm]{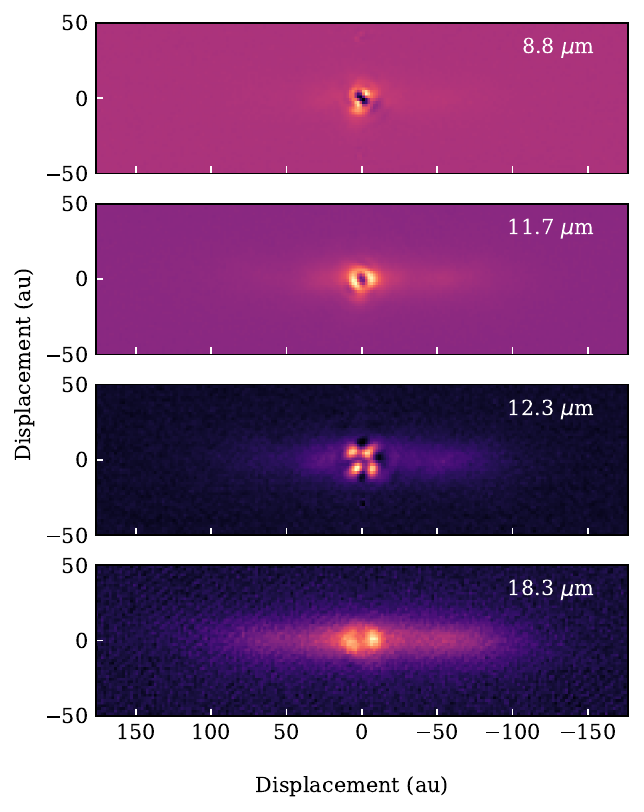}
    \caption{Central component-subtracted images using best-fit flux values shown in Table~\ref{tab:starflux}, where the central component represents the stellar flux plus any potential unresolved inner warm dust. These images suggest the presence of a potentially resolved dust population at the inner regions of the system. }
    \label{fig:star_subtraction}
\end{figure}

\subsection{Radial profile}
\label{sec:rp}
The multi-wavelength images of $\beta$~Pic all provide information on its radial structure despite its edge-on viewing perspective. To best leverage this information and derive realistic constraints while minimising model assumptions, we applied the \texttt{rave} algorithm \citep{Han2022, Han2025} to recover the face-on radial surface brightness profile at each wavelength. \texttt{rave} fits the de-projected and deconvolved radial profile of optically-thin disks non-parametrically, i.e., without assuming a functional form parametrised by a number of functional parameters. 
This is advantageous given that such an approach mitigates against biases towards an assumed shape of the radial profile and provides realistic uncertainties based on the range of possible models that is independent of assumptions on the functional form. Importantly, \texttt{rave} recovers the radial profile independent of height and inclination assumptions and can be applied even if the disk is perfectly edge-on. This is achieved by fitting to the flux distribution summed onto the disk's major axis, which is an observable that does not vary with the vertical distribution or projection angle. The independence of height assumptions enables these radial profiles to be used when modelling the vertical structure in Section~\ref{sec:vertical}. 

The algorithm \texttt{rave} operates under the assumption that the disk is azimuthally symmetric and optically thin, such that discrete axisymmetric annuli can be fitted to the vertically summed emission profile to find a flux distribution among the annuli that reproduces this profile. Although the disk is known to be vertically warped, its effect on the radial profile is negligible as the image is vertically summed over before being fitted by \texttt{rave}. The emission of the bright central point source in the disk, which encompasses emission from the star, is fitted to in this procedure by placing an unresolved central annulus in the fit. Treating the SW clump clearly visible in the mid-infrared as a separate component from the underlying disk, we assumed the NE half of the disk to be representative of the underlying disk emission upon which the SW clump is super-imposed. We therefore only fitted \texttt{rave} to recover the radial profile and central component's flux density based on the NE emission of all mid-infrared images. 
At ALMA wavelengths, however, this asymmetry is all but negligible given their low significance as discussed in Section.~\ref{sec:asymmetries}. 
We therefore treated the disk as axisymmetric for the purpose of obtaining ALMA radial profiles with \texttt{rave}. 

\begin{table}
    \centering
    \caption{Fitted unresolved central component flux densities. }
    \label{tab:starflux}
    \begin{tabular}{ll}
    \hline \hline
    Wavelength ($\mu$m) & Central component flux (Jy) \\ \hline
    8.8        & $2.35^{+0.03}_{-0.13}$         \\
    11.7       & $1.66^{+0.04}_{-0.10}$         \\
    12.3       & $1.55^{+0.04}_{-0.15}$         \\
    18.3       & $1.15^{+0.04}_{-0.08}$         \\
    24.5       & $1.23^{+0.14}_{-0.13}$         \\
    887        & $1.5^{+2.6}_{-1.0} \times 10^{-4}$     \\
    1300       & $5^{+4}_{-3} \times 10^{-5}$  \\ \hline
    \end{tabular}
\end{table}

The recovered radial profiles of all observations in this study are shown in Fig.~\ref{fig:radial_profiles}, with any relevant hyperparameter values used for the fitting described in the figure caption, and the fitted central component flux densities (which are not part of the radial profiles displayed) are listed in Table~\ref{tab:starflux}.

In the millimeter, the Band 6 (0.9\,mm) and 7 (1.3\,mm) radial profiles show qualitatively consistent features as expected for two closely positioned wavelengths. The millimeter emission peaks at a radius of 70 to 110\,au, with falling emission out to at least 200 au. The Band 7 radial profile suggests there to be a change in steepness on the outer edge at $\sim$150\,au. The surface brightness of the section of outer edge interior to 150\,au appears to be largely symmetric with the inner edge, whereas the region exterior to 150\,au drops off more slowly. 

The mid-infrared profiles generally reveal a steady drop-off from the star. Emission increases at every disk location from 8.8 towards 24.5 microns. No robust substructures are detected at the sensitivity and resolution of these observations. 
Note that the mid-infrared radial profiles do not cover the region closer to the star since PSF variations and asymmetries introduce large uncertainties in the disk emission close to the bright stellar core. 
To determine the extent of the uncertain radial regions, we plot in Fig.~\ref{fig:1D_residuals} the one-dimensionalised flux of each image and that reproduced by the fitted model (vertically integrated out to 44\,au from the major axis in the mid-infrared and 50\,au in the millimeter), which acts as a diagnostic for assessing the quality of the fit. We examined the fit to the emission at different distances from the star and only displayed the radial profile corresponding to regions away from the PSF of the bright stellar core where the disk emission is well-reproduced by the fit as suggested in Fig.~\ref{fig:1D_residuals}. 

We find some evidence of a marginally resolved inner dust component in the inner regions of the system.
The radial profile modelling with \texttt{rave} fitted the stellar flux plus any potential unresolved inner dust, which we have collectively referred to as the ``central component''. 
Fig.~\ref{fig:star_subtraction} displays the mid-infrared images with the PSF subtracted, where the subtracted PSF was normalised to the flux densities listed in Table~\ref{tab:starflux}. While the distribution of any emission in the inner region is too heavily affected by PSF subtraction artefacts to accurately model (giving the circularly asymmetric patterns in Fig.~\ref{fig:star_subtraction}), these PSF-subtracted images appear to show residual close-in emission in the form of a wider bright core than the reference star, which could arise from a marginally resolved inner dust component in the inner regions of the system. 
JWST/MIRI MRS observations have also suggested the presence of an inner dust component as close-in as $<$7\,au from the star \citep{Worthen2024}. 
However, as the inner regions are beyond the reliable radial profile interval, we do not study it further in our modelling, which is primarily concerned with the outer disk. 

\subsection{Spatially resolved spectral energy distribution}
\label{sec:sed}

\begin{figure*}
    \centering
    \includegraphics[width=8.5cm]{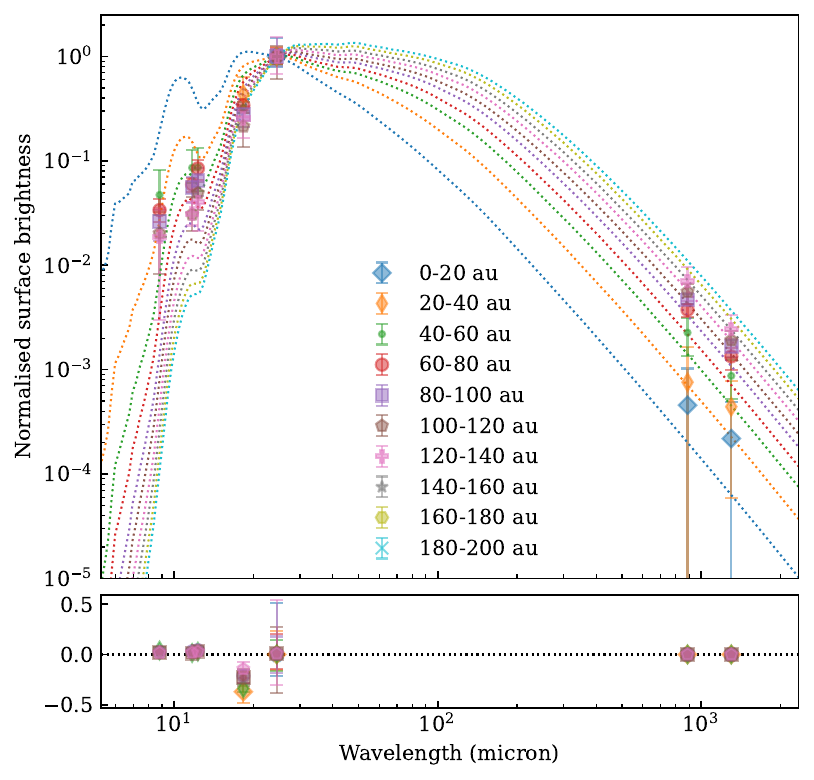}
    \includegraphics[width=8.5cm]{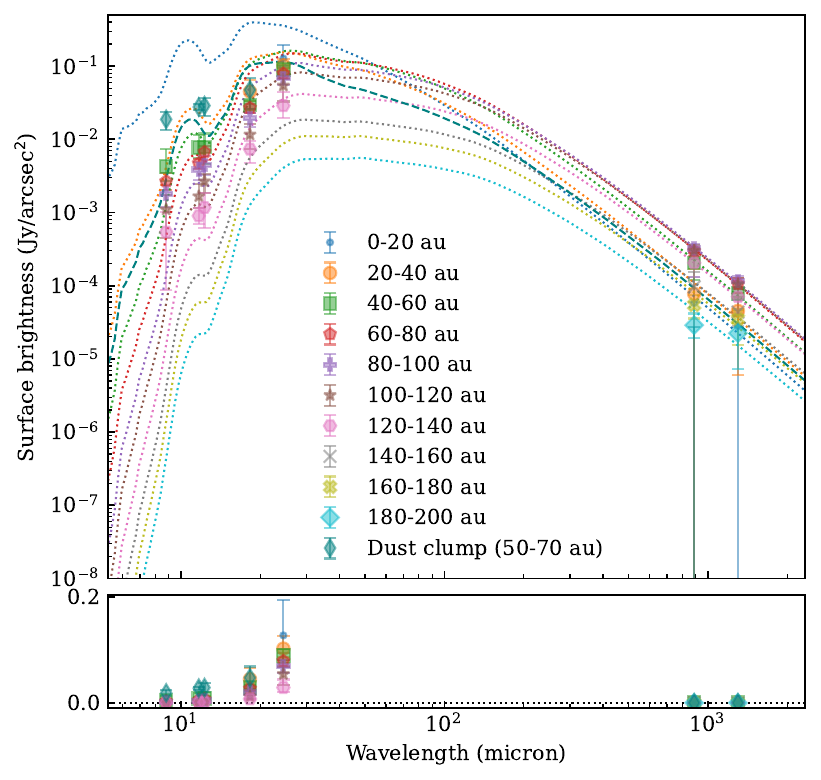}
    \caption{The spatially resolved SED of different radial regions in $\beta$~Pic's debris disk. Residuals are displayed on the bottom panels. Flux values for each radial region are taken from the radial profiles shown in Fig.~\ref{fig:radial_profiles}. The panel on the left is normalised to the 24.5~$\mu$m flux to emphasise the mid-infrared and mid-infrared to millimeter spectral index, whereas the panel on the right is not. Dotted lines show SED models for each radial region. The SW dust clump is plotted as a separate region in the right panel, whereas the remaining mid-infrared points only represent the NE side of the disk without the SW dust clump. Note that while the SED models are plotted over the entire wavelength range for all radial regions, some wavelength and radial region combinations do not have corresponding flux measurements due to limitations in the reliable interval of the radial profiles. }
    \label{fig:SED}
\end{figure*}

\begin{figure*}
    \centering
    \includegraphics[width=17cm]{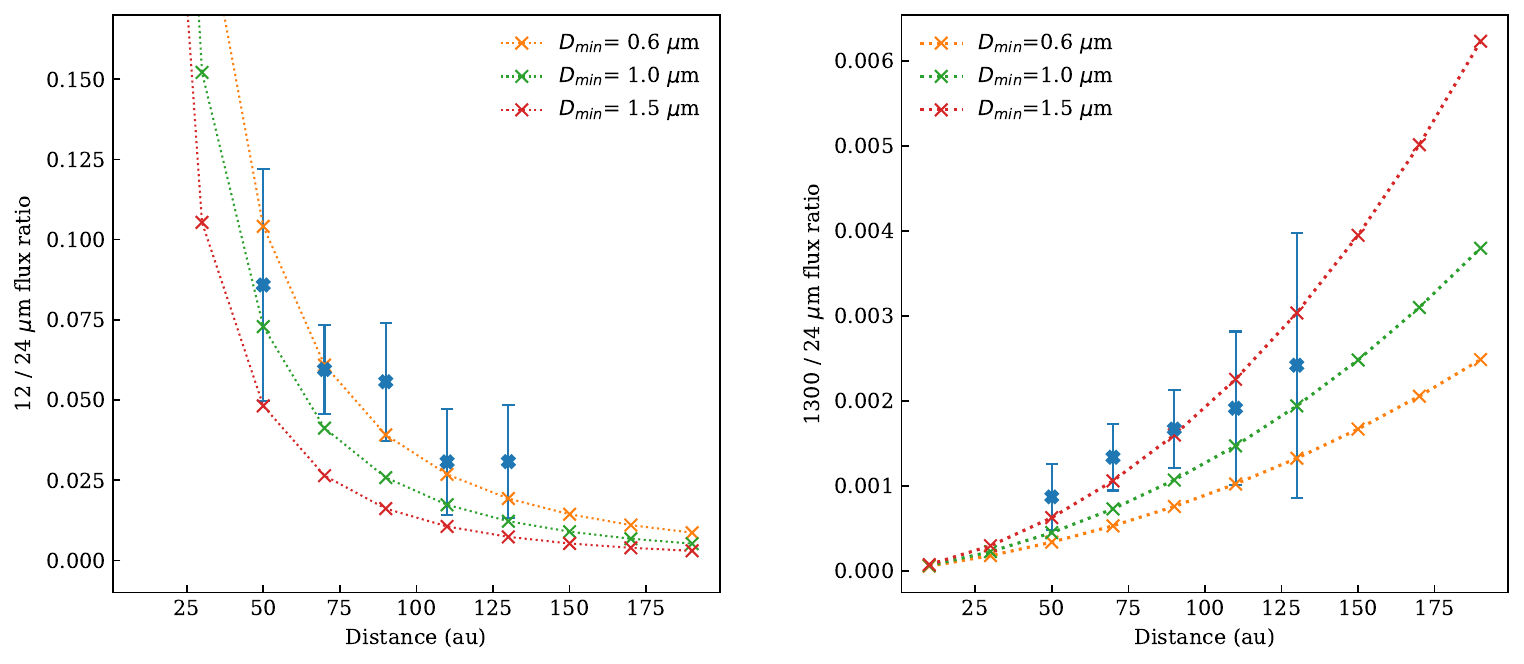}
    \caption{The 11.7 to 24.5 and 1300 to 24.5~$\mu$m flux ratios across different radial regions in the NE side of the disk. The spectral index values are derived from the SEDs shown in Fig.~\ref{fig:SED}. Each dotted line corresponds to an SED model with a different minimum dust grain size in the disk (as labelled) assuming asteroidal composition. }
    \label{fig:spectral_index}
\end{figure*}

\begin{figure}
    \centering
    \includegraphics[width=8cm]{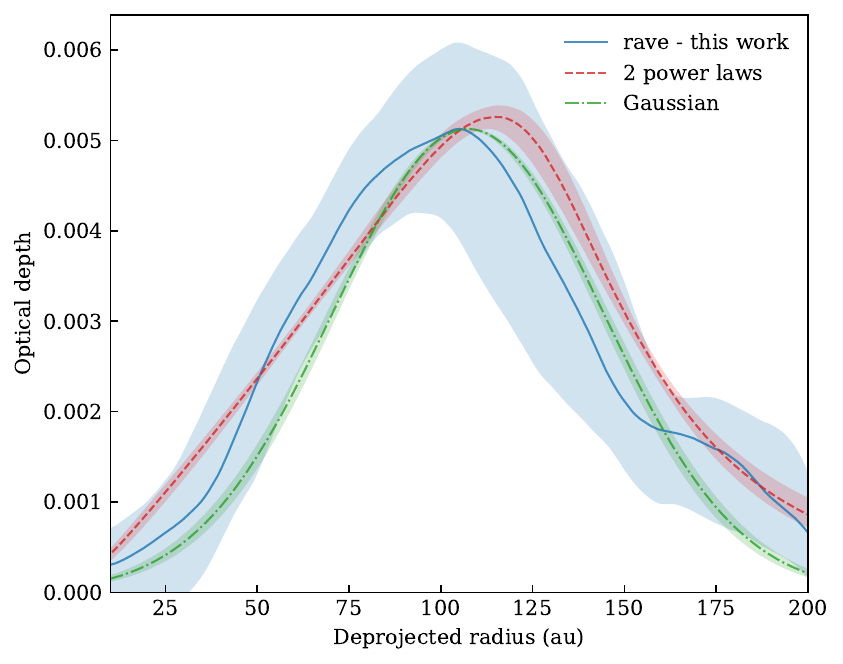}
    \caption{The radial optical depth profile of $\beta$~Pic's debris disk. This is the mean profile derived from the ALMA Band 6 and 7 radial surface brightness profiles in Fig.~\ref{fig:radial_profiles} assuming asteroidal composition and a minimum grain size of 1~$\mu$m. A surface density profile parameterised by two power laws fitted in \citet{Han2026b} and a Gaussian surface density profile fitted in \citet[Model~3 in that study]{Matra2019} are plotted for comparison, with both normalised to the peak optical depth in this work.} 
    \label{fig:surface_density}
\end{figure}

The multi-wavelength radial profiles which we have derived enable the construction of SEDs for different radial regions in the disk to study its dust grain properties. We partitioned the disk between 0 to 200\,au from the star into radial bins that are each 20\,au wide, which is equal to the resolution of the lowest-resolution image in ALMA Band 7. By obtaining flux measurements from the radial profiles in Fig.~\ref{fig:radial_profiles} averaged over each radial bin, we assembled the spatially resolved mid-infrared to millimeter SED of the disk as shown in Fig.~\ref{fig:SED}. The uncertainties were determined using the extreme values within the range of possible models for the radial profile. 
For some of the radial bins we omit some of the mid-infrared wavelengths. This is because the mid-infrared radial profiles are unreliable too close to the star (see Section~\ref{sec:rp} and Fig.~\ref{fig:1D_residuals}). Additionally, none of the mid-infrared profiles are reliable beyond 150 au, and including them in those radial bins would bias the SED analysis in log space.

To better visualise the differences in spectral indices at different radii, we also plotted the SEDs normalised to the 24.5~$\mu$m flux in the left panel of Fig.~\ref{fig:SED}, as well as the flux ratios between 11.7 and 24.5~$\mu$m and between 1300 and 24.5~$\mu$m in Fig.~\ref{fig:spectral_index}. We observe that in the mid-infrared regime, the spectral index steepens when moving further out in the disk, while the reverse is true for the spectral index between 24.5~$\mu$m and the millimeter wavelength regime. 

To test whether this trend can be explained by the radial temperature distribution with a single grain type and size distribution of dust, we modelled the emission of the disk as
\begin{equation}
    \frac{d\xi}{dD} = Q_\text{abs}(\lambda, D) B_\nu(T(r, D), \lambda) \frac{d\tau(r, D)}{dD},
\end{equation}
\noindent where ${d\xi}/{dD}$ is the surface brightness per grain size bin at any given radius and wavelength, $Q_\text{abs}$ is the absorption efficiency which reflects optical properties resulting from the physical and chemical composition of the dust grains, $B_\nu$ is the Planck Function which carries the effect of the dust temperature, $T$, and $\tau$ is the geometric optical depth which describes the spatial and size distribution of material in the disk. These quantities are expressed as functions of $r$, the radius from the star, $\lambda$, the wavelength, and $D$, the diameter of the dust grain. 

Assuming a stellar luminosity of 8.52~L$_\odot$ and effective temperature of 8190~K, we simulated the optical properties and temperature profile of the dust grains following the approach in \citet{Wyatt2002} using the using the \texttt{astrodust\_optprops} code \citep{Sommer2025}, calculating $Q_\text{abs}$ based on Mie theory or Rayleigh-Gans theory depending on the relative size of the dust grains and wavelength. The refractory indices of the dust grains used in the calculation were obtained by combining those of its constituent material \citep{Li1997, Li1998} using Maxwell-Garnett Effective Medium Theory \citep{Bohren1983}, which we assumed to consist of a silicate core occupying $1/3$ of the volume of each dust grain and an amorphous organic refractory mantle occupying the rest, analogous to asteroidal compositions. 

To integrate ${d\xi}/{dD}$ over grain size and model the normalised surface brightness profile $\xi/\tau$, we assumed a grain size distribution described by
\begin{equation}
    \frac{dN}{dD} \propto D^{-3.5}
\end{equation}
\noindent between $D_\text{min}$ and $D_\text{max}$ and 0 elsewhere, consistent with a steady-state collisional cascade grain size distribution in which grains smaller than a certain limit are expected to be removed from the system by stellar radiation pressure \citep{Wyatt2008}. The power law exponent of -3.5 is consistent with the value of -3.49\,$\pm$\,0.08 measured by \citet{MacGregor2016} based on the unresolved SED of $\beta$~Pic. $\xi/\tau$ is not sensitive to the value of $D_\text{max}$ which we set to be 1~m, as long as $D_\text{max}$ is much greater than the maximum wavelength of interest, which is $\sim$1~mm in this study. 

However, $\xi/\tau$ is sensitive to $D_\text{min}$, and we plotted the flux ratios for various values of $D_\text{min}$ in Fig.~\ref{fig:spectral_index} to compare with the observationally derived values. We find that a minimum grain size of 0.6~$\mu$m is able to reproduce the mid-infrared spectral indices, whereas a slightly larger minimum grain size of 1.5~$\mu$m is preferred to best reproduce the mid-infrared to millimeter spectral indices. However, a minimum grain size of 1.0~$\mu$m appears to provide a compromise that broadly agrees with both sets of spectral indices given the relatively large uncertainties in these measurements, and is able to reasonably reproduce the normalised SED in the left panel of Fig.~\ref{fig:SED}. 

The optical properties in the model also predict the radiation pressure blowout size for the dust grains, $D_\text{bl}$. Assuming a stellar mass of 1.75~M$_\odot$, the dust composition assumed corresponds to a $D_\text{bl}$ value of 5.4~$\mu$m, which is significantly larger than the minimum grain size of 1.0~$\mu$m required to reproduce the SED with our model. 
This $D_\text{min}$/$D_\text{bl}$ of $\sim$0.2 sits within the typical range spanned by A stars, though near the lower end of that range \citep{Marshall2026}.
Future studies may wish to test the extent to which $D_\text{min}$ is affected by assumptions in the grain composition or in the size distribution from the -3.5 power law index assumed here. 

Assuming $D_\text{min} = 1.0\,\mu$m, we used the modelled normalised surface brightness, $\xi/\tau$, and the observed surface brightness profiles of ALMA Bands 6 and 7 which reliably cover the full radial range from 0 to 200\,au to infer the optical depth, $\tau(r)$. The mean radial optical depth profile is shown in Fig.~\ref{fig:surface_density}. Using this profile for $\tau(r)$, we derived models for $\xi(\lambda, r)$ and hence the full SED models (without normalising $\tau$) as shown in the right panel of Fig.~\ref{fig:SED}. 

In Fig.~\ref{fig:SED}, we also plotted the SED of the SW dust clump measured by subtracting off emission from the opposite point about the star in the NE side of the disk. The peak region was defined as the rectangle between 50 and 70\,au in projected separation along the major axis and vertically within 10\,au of the major axis. The clump is not detected at 24.5~$\mu$m due to the low S/N, preventing the 12/24 and 1300/24~$\mu$m flux ratio from being calculated. However, its 12/18~$\mu$m flux ratio of $0.4^{+0.3}_{-0.2}$ and 887/18~$\mu$m flux ratio of $<2\times10^{-3}$ are significantly different from the underlying disk, for which the values are $0.12^{+0.06}_{-0.05}$ and $(8 \pm 2) \times 10^{-3}$ respectively. The clump's SED can be fitted with a surface density equal to one third of the corresponding NE region but requires a lower minimum grain size of $\sim$0.2~$\mu$m. Such a dust population would be significantly smaller than both the blowout size and $D_\text{min}$ in the rest of the disk, and confirms the finding of the dust clump's SED being unique from the rest of the disk in \citet{Telesco2005}. 

\section{Vertical structure}
\label{sec:vertical}

\begin{figure*}
    \centering
    \includegraphics[width=17cm]{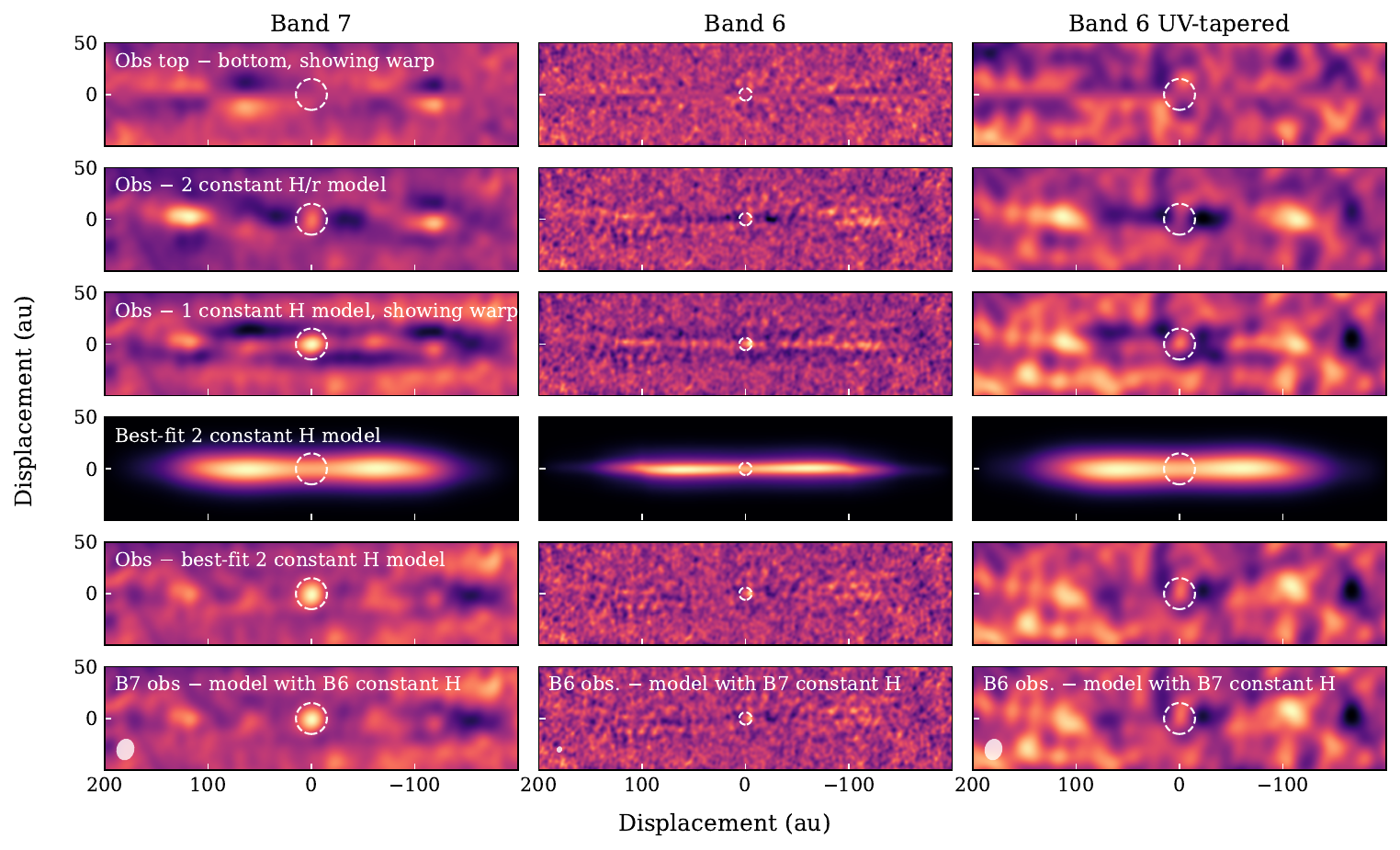}
    \caption{Gallery of plots used to explore the millimeter vertical structure of $\beta$~Pic referenced throughout Section~\ref{sec:vertical}. From left to right, the columns represent plots for ALMA Band 7, Band 6 and Band 6 again but UV-tapered to the resolution as the Band 7 observations. Each row represents a different type of plot. The top three rows correspond to exploratory plots which help better understand the vertical features present in the disk, whereas the bottom three rows correspond to the full fitted model accounting for these features. From top to bottom, these are (1) emission above the major axis minus emission below the major axis, with residuals suggestive of a vertical warp detected in these millimeter observations; (2) residuals from subtracting the best-fit model containing two vertical populations presented in \citet{Matra2019}, each with a constant aspect ratio ($H/r$), suggesting that constant aspect ratio models may not fully explain the millimeter disk structure; (3) residuals from subtracting a flat midplane model (i.e., no warping), suggesting that a closer fitting model will need to account for the vertical warp at the sensitivity of these observations; (4) best-fit model accounting for the warp with two constant height ($H$) Gaussian components, fitted independently for Band 7 and 6; (5) residuals of the best-fit models and (6) comparing the independent Band 7 and 6 vertical profile models by subtracting the model for one band from the observations of the opposite band (using the appropriate radial profiles and convolving the model the with the appropriate beam to achieve identical resolutions). Emission from the star is labelled with white dashed circles. }
    \label{fig:height_plots}
\end{figure*}

\begin{figure}
    \centering
    \includegraphics[width=8.5cm]{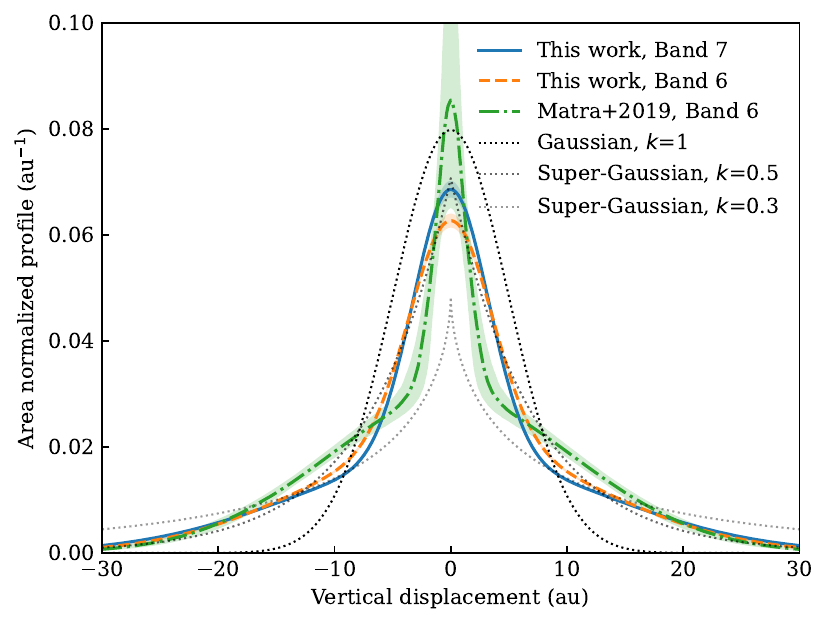}
    \caption{Comparison of fitted vertical profiles. The best-fit vertical profiles are shown for ALMA Band 7 and 6 with the two constant $H$ model in this study and the two constant $H/r$ model in \citet{Matra2019}. A few super-Gaussian functions with a standard deviation of 10\,au are plotted for comparison, which are modified Gaussian functions with the magnitude of the exponent raised to a power of $k$, and has been used to parametrise vertical profiles in debris disks in other studies. All profiles are normalized by the area under the curve. }
    \label{fig:vertical_profiles}
\end{figure}

We begin this section by exploring vertical substructures present in ALMA observations before fitting a model that accounts for these substructures to measure the millimeter scale height. We then apply this model to measure the mid-infrared scale height and compare its value across wavelength. 

\begin{figure*}
    \centering
    \includegraphics[width=17cm]{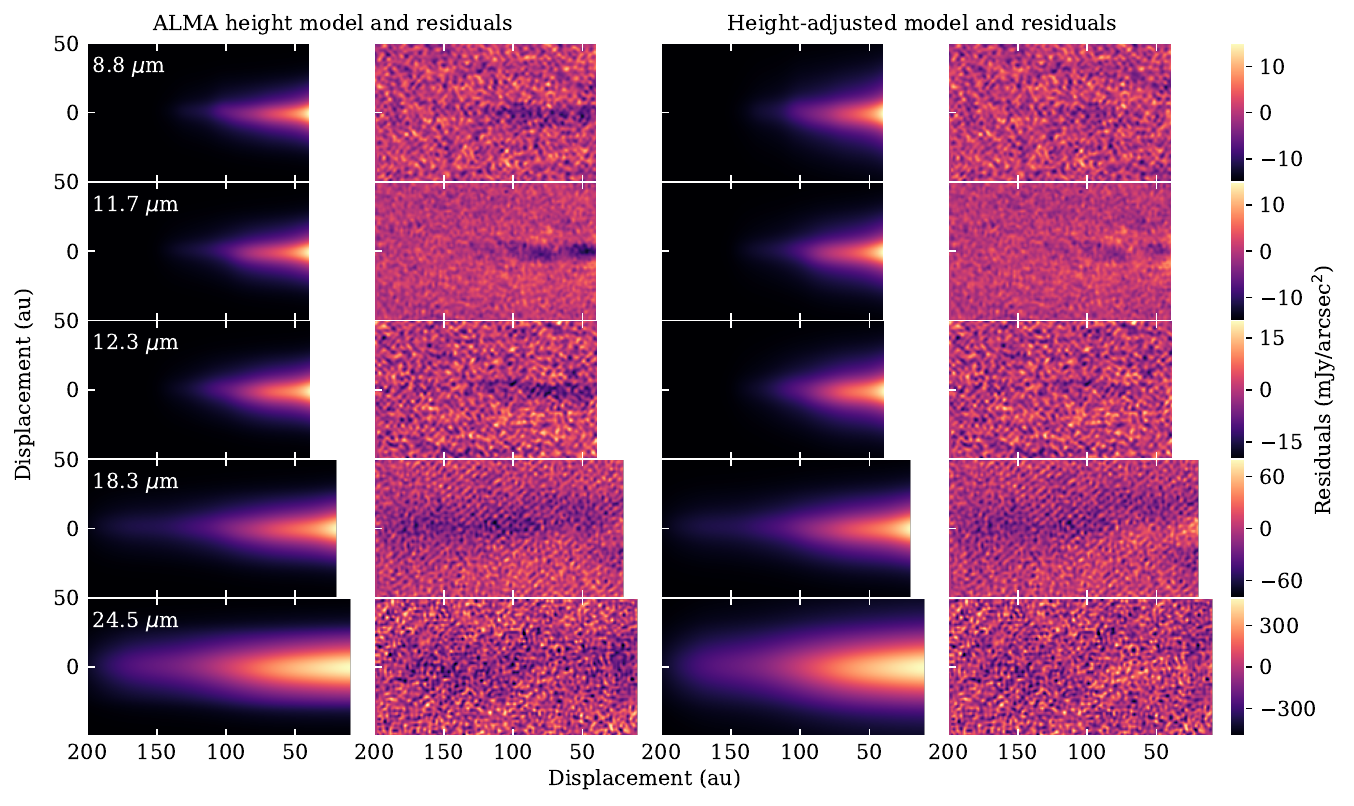}
    \caption{Scale height models for the NE arm of the disk in the mid-infrared. The ``ALMA height models'' on the left show models generated using the radial profile and PSF at the corresponding mid-infrared wavelength but with the vertical structure model fitted to ALMA Band~7. The over-subtracted midplanes in these residuals indicate that the mid-infrared scale height should be significantly larger than at ALMA wavelengths. The ''height-adjusted models'' on the right show adjustments to these models by keeping the shape of the ALMA vertical profile and the structure of the warp but increasing the height by a factor with its best-fit value shown in Table.~\ref{tab:MCMC}. }
    \label{fig:mir_models_res}
\end{figure*}

\begin{figure}
    \centering
    \includegraphics[width=8.5cm]{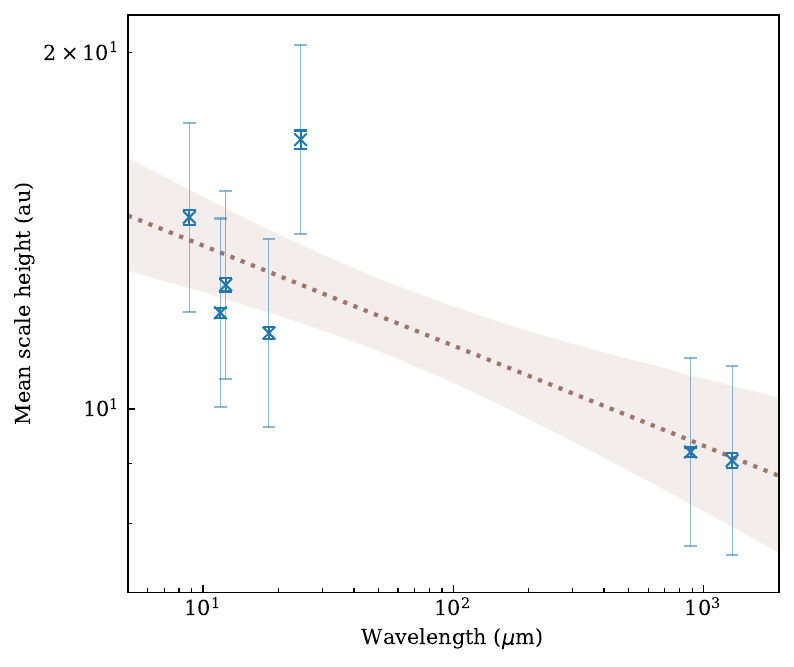}
    \caption{The representative scale height of the debris disk of $\beta$~Pic across mid-infrared and millimeter wavelengths. The representative scale height is the weighted mean between the two height populations in the fitted model at each wavelength as displayed in Table~\ref{tab:MCMC}. Each point has two sets of uncertainties. The darker, thicker error bars indicate those returned from the MCMC fit of the two $H$ model to the observations, whereas the lighter, thinner error bars indicate the uncertainties estimated by fitting a power law to these best-fit height values as a function of wavelength, which was also fitted with MCMC. The dotted line and shaded region indicate the fitted median and 1$\sigma$ power law scaling between scale height and wavelength. }
    \label{fig:height}
\end{figure}

\subsection{Vertical warp at millimeter wavelengths}
\label{sec:almaheight}

Scattered light observations of $\beta$~Pic \citep{Heap2000, Golimowski2006, Apai2015} have shown the disk to be vertically warped in such a way that the emission is tilted counterclockwise from the major axis on both arms of the disk at approximately the radius of the SW dust clump. No significant vertical substructures have been found from ground-based mid-infrared observations owing to its poorer resolution and sensitivity, whereas JWST/MIRI imaging has revealed the extended and tilted secondary disk to the NE and ``cat's tail'' to the SW \citep{Rebollido2024} as discussed in Section.~\ref{sec:asymmetries}. 

In the millimeter, dust continuum modelling \citep{Matra2019} based on the ALMA Band 6 image also found no obvious warping of the disk by fitting Gaussian profiles to vertical slices along the major axis, however the ALMA CO distribution \citep{Dent2014, Matra2017} was found to be asymmetric in a way analogous to the mid-infrared dust clump and tilted in a way consistent with the scattered light warp. \citet{Matra2019} also found that the vertical distribution of material in the disk appears to be non-Gaussian. The vertical profile was better fit by the superposition of two Gaussians, each with a constant aspect ratio ($h = H/r$), which the study suggested could reflect the presence of two dynamical populations, with one having a higher inclination (and therefore likely eccentricity) dispersion than the other. 

Taking the radial profiles derived with \texttt{rave} in Section~\ref{sec:rp}, we constructed a model with this two-Gaussian component height profile using the best-fit position angle and vertical profile assuming constant aspect ratio from \citet[Model 3]{Matra2019}. We assumed the inclination to be 89$^\circ$ in this and subsequent models, informed by the best-fitting inclinations in the \citet{Matra2019} model. The residuals in both bands are shown in Fig.~\ref{fig:height_plots} row~2. The close resemblance between the radial profile which we derived with \texttt{rave} and the parametric Gaussian radial profile in \citet[Fig.~\ref{fig:surface_density}]{Matra2019} should result in no substantial difference between this and the \citet{Matra2019} model at Band 6, as confirmed by the similar residual structures compared to Fig.~6 in \citet{Matra2019}. However, significant residuals appear in the Band 7 data, most notably under-subtracted lobes at 120\,au on both sides of the disk surrounded by over-subtracted regions, as well as over-subtracted inner regions of the disk. This is also seen in the Band 6 data when UV-tapered to the same resolution as Band 7 in Fig.~\ref{fig:height_plots}. Furthermore, the residual features near the two lobes appear to be slanted from the midplane in opposite directions but consistent with each other and across the two bands, suggesting that the millimeter disk displays warp in projection. The same residual structures remain if a slightly different inclination were to be assumed, such as 87.7$^\circ$ as fitted by \citet{Zawadzki2026}. 

To validate the presence of a vertical warp, we subtracted the emission below the major axis of the disk from the emission above assuming a position angle of 29.7$^\circ$, as shown in Fig.~\ref{fig:height_plots} row~1. A checkerboard pattern emerges in the self-subtracted images, as expected from a vertically warped midplane, which is again seen in both bands but is particularly pronounced in Band~7. 

To further verify the presence of a warp, we subtracted a symmetric ``test model'' (rather than self-subtracting the disk image) in which the vertical profile is a single Gaussian with a constant scale height throughout the entire disk. The radial profile remains as the fitted \texttt{rave} profile, noting that the \texttt{rave} radial profile is derived independently from any height or inclination assumptions. Subtraction of such a model with a scale height of 9\,au from the two bands is shown in Fig.~\ref{fig:height_plots} row~3. 

Two observations follow from the panels in this row. Firstly, the residuals demonstrate a clear warp in Band 7. The presence of any warping is less apparent from the residuals in Band 6, but adjusting for its different resolution reveals hints of such a warp. These together suggest that $\beta$~Pic’s debris disk is warped in the millimeter too. The peak vertical displacement in the warp appears to occur at a projected separation of $\sim$70\,au from the star, which coincides with the known vertical warp at optical wavelengths \citep{Apai2015}. 

Secondly, although the constant, single-Gaussian height model is clearly not (and was not intended to be) a good fit for the height profile, it manages to produce a largely constant (albeit warped) banded structure in the residuals. This is unlike the residual structure when assuming a scale height proportional to radius, in which the residual structure varies across the disk, such as the two residual lobes in Fig.~\ref{fig:height_plots} row~2. This suggests that the radial variations in the vertical height could be better described as being constant under our inclination assumption, rather than being proportional to radius. 

The two observations together point to the fact that an improved model in thermal emission should account for both the vertical warping in the disk and the more constant scale height across radius. We develop such a model in the following section. 

\subsection{The scale height in ALMA observations}
\label{sec:almamodel}
We now construct a model that accounts for (1) the vertical warp, (2) the relatively constant scale height across radius and (3) the non-Gaussian vertical distribution as discussed in the previous section. Rather than fitting a large number of model parameters and attempting to accurately constrain all vertical substructures simultaneously, our aim is simply to achieve a model with reasonable residuals such that our primary quantity of interest, the vertical height, can be derived consistently and compared across wavelength. 

To account for the vertical warping, we partition the disk into 2 radial regions in the orbital plane with different position angles, but the same inclination of 89$^\circ$, such that they tilt in opposite directions relative to each other's midplane to create the warp. Based on the residual structures in Fig.~\ref{fig:height_plots} row~3, we infer that the warp can be reproduced when the two radial regions transition at a radius of 100\,au from a position angle of -1.2$^\circ$ to 0.5$^\circ$ relative to the overall position angle of 29.7$^\circ$ determined from the observations, resulting in straight residual structures that lie along the major axis without warping. Note that the transition radius for the position angle does not have to match the location of peak displacement of the warp, since the peak location is determined by the combined effect of the two radial components as well as effects due to the surface brightness profile. In practice we find by trial and error that under this model construction, the reasonable range of transition radius is within $\sim$10\%. The representative scale height that we subsequently fit does not sensitively depend on the specific choice of transition radius within this small range and variations as a result of any dependence is within uncertainties. Assuming a less inclined viewing angle would decrease the subsequent best-fit scale height, however the trend in any scale height variations with wavelength which we aim to investigate is expected to be preserved. 

To account for the non-Gaussian vertical distribution, we parametrised the vertical distribution to be the sum of two Gaussian components with standard deviations of $H_1$ and $H_2$ and with a fractional flux of $f_1$ in $H_1$ that is independent of radius. To account for the relatively constant scale height across radius, $H_1$ and $H_2$ are also assumed to be constant across the disk. As the radial profile is already derived non-parametrically with \texttt{rave} and is independent of height assumptions, this leaves only 3 free parameters in this model that together describe the scale height of the disk at any wavelength. 

We used a Markov chain Monte Carlo (MCMC) algorithm to sample the $(H_1, H_2, f_1)$ parameter space independently for Bands 7 and 6, maximising the log-likelihood computed from the squared difference between the model image (convolved with the beam) and observations. The MCMCs were carried out with the \texttt{emcee} package \citep{emcee}, leaving the uncertainties in the image as a free parameter and enforcing $H_1 < H_2$ so that $H_1$ represents the vertically thinner component. Due to uncertainties in the fitted stellar flux in Section~\ref{sec:radial} and to avoid these uncertainties biasing the vertical height fitting, the region around the star as labelled with white dashed circles in Fig.~\ref{fig:height_plots} was not used to compute the squared difference during model fitting. 

The fitted parameters are displayed in Table~\ref{tab:MCMC} and the residuals in Fig.~\ref{fig:height_plots} row~5. The residual structures are significantly improved compared to the double vertical component with constant aspect ratio model in Fig.~\ref{fig:height_plots} row~2. The residual lobes on either side of the disk are now significantly reduced in both bands, and any potentially identifiable residual features lie along the disk's midplane without significant warping. Some of these potential residual features could relate to the asymmetric features shown in Fig.~\ref{fig:image180_alma} and discussed in Section~\ref{sec:asymmetries}, which are intrinsic to the disk. While we assumed the inclination to be 89$^\circ$, the scale height does not sensitively depend on the inclination assumption within a $\sim$1$^\circ$ interval, which contains the uncertainty range in \citet{Matra2019}. Overall, we consider this model to be satisfactory for the purpose of measuring the scale height in the ALMA observations. 

\begin{table*}
    \centering
    \caption{Scale heights fitted to the observations across all wavelengths. }
    \label{tab:MCMC}
    \begin{tabular}{lllllll}
    \hline\hline
                 & Wavelength ($\mu$m) & $H_1$ (au)             & $H_2$ (au)              & $f_1$                     & $\phi_\text{B7}$       & Representative scale height (au) \\ \hline
    Millimeter   & 1300 (ALMA Band 6)  & $3.49^{+0.09}_{-0.07}$ & $12.31 \pm 0.14$        & $0.37 \pm 0.01$           &                        & $9.05 \pm 0.13$              \\
                 & 887 (ALMA Band 7)   & $3.20 \pm 0.09$        & $13.33 \pm 0.07$        & $0.408 \pm 0.006$         &                        & $9.20 \pm 0.08$              \\
    Mid-infrared & 24.5                &                        &                         &                           & $1.88^{+0.03}_{-0.04}$ &  $14.5 \pm 0.2$              \\
                 & 18.3                &                        &                         &                           & $1.29^{+0.06}_{-0.01}$ &  $12.06 \pm 0.12$            \\
                 & 12.3                &                        &                         &                           & $1.41^{+0.01}_{-0.01}$ &  $12.73 \pm 0.16$            \\
                 & 11.7                &                        &                         &                           & $1.32^{+0.03}_{-0.00}$ &  $11.59 \pm 0.13$            \\
                 & 8.8                 &                        &                         &                           & $1.61^{+0.02}_{-0.03}$ &  $16.9 \pm 0.3$              \\ \hline
    \end{tabular}
    \tablefoot{The millimeter vertical profiles were parametrised by the sum of two Gaussians, with a fraction $f_1$ of the flux in $H_1$ and the remainder in $H_2$, both of which are constant across radius. The mid-infrared scale heights were only fitted with one free parameter, $\phi_\text{B7}$, which represents vertical height as a multiple of ALMA Band 7 (887~$\mu$m) while retaining the shape of the fitted Band 7 vertical profile. The representative scale height ($\bar{H}$) corresponds to the mean of $H_1$ and $H_2$ weighted by $f_1$ and $(1 - f_1)$, with uncertainties estimated using MCMC by assuming that there exists a power-law scaling between scale height and wavelength and that the uncertainties are identical in log space.}
\end{table*}

Fig.~\ref{fig:vertical_profiles} shows the best-fit vertical profiles, which is constant throughout the disk. The best-fit profiles under the two constant aspect ratio model in \citet{Matra2019} is plotted at a radius of 100\,au for comparison, which is approximately where the surface brightness profile peaks. While specific values relating to the two height components differ in Table~\ref{tab:MCMC}, all three fitted vertical profiles in Fig.~\ref{fig:vertical_profiles} appear to be broadly consistent, with the Band~6 vertical profile from \citet{Matra2019} showing a slightly pointier peak. 

For ease of comparison, we may define a ``representative scale height'' for the 2-component model as the weighted mean between the height of the 2 components, 
\begin{equation}
    \bar{H} = f_1 H_1 + (1 - f_1) H_2.
\end{equation}
The representative heights at the two bands are displayed in Table~\ref{tab:MCMC}, which are consistent within uncertainties. 

To investigate whether the subtle difference in the shape of the vertical profile between Bands 6 and 7 could be real, we produced an alternative model for Band~7 using the Band~7 radial profile and effective beam but with the best-fit vertical profile from Band~6, and vice versa. The residuals of these ``hybrid models'' are shown in Fig.~\ref{fig:height_plots} row~6 and are structurally consistent with the best-fit models in row~5, suggesting the slight difference in their vertical profile shapes are insignificant.

\subsection{Measuring the mid-infrared scale height using the ALMA ruler}
\label{sec:mirheight}

In this section, we aim to constrain the scale height of the disk in the mid-infrared. 
To test whether the ALMA vertical height may also explain the mid-infrared observations, we simulated mid-infrared model images based on the radial profiles at each of the 5 mid-infrared wavelengths and the corresponding PSFs, but using the best-fit vertical profile and warping for ALMA Band~7 (which is largely consistent with Band~6). While the mid-infrared and millimeter images were previously assumed to be at two different position angles to align the major axis horizontally for display purposes, a consistent vertical height comparison in this section requires that all images be aligned to the same angle. Assuming negligible uncertainties in the telescopes' rotation angle during pointings, we assumed the position angle of the disk to be 29.7$^\circ$ at all wavelengths, consistent with the value determined in \citet{Matra2019}. 

The resulting models and residuals are shown in the left columns of Fig.~\ref{fig:mir_models_res}. The significantly over-subtracted midplanes in the residual images clearly indicate that the best-fit ALMA vertical structure underestimates the height at all mid-infrared wavelengths, suggesting that the mid-infrared scale height is larger than in the millimeter. 

While the mid-infrared observations also provide important information on the vertical structure, they lack the sensitivity to reveal the functional form of the vertical profile in the way that the ALMA images do through our analysis in Section~\ref{sec:almaheight}. 
To quantify the mid-infrared scale height consistently for comparison with ALMA, we took a simpler approach by fixing the shape of the vertical profile and the vertical warping to be the same as ALMA Band 7, fitting only to the width of this distribution. We characterised the mid-infrared height as a multiple of ALMA Band~7, $\phi_\text{B7}$, fixing $f_1$ at the fitted value for Band~7 and setting
\begin{align}
    H_{1, \text{mid-IR}} &= \phi_\text{B7} \, H_{1, \text{B7}}\\
    H_{2, \text{mid-IR}} &= \phi_\text{B7} \, H_{2, \text{B7}}.
\end{align}

$\phi_\text{B7}$ is therefore the only free parameter of interest. We sampled $\phi_\text{B7}$ independently for each of the 5 mid-infrared wavelengths with MCMC, including uncertainties in the image as a free parameter. The best-fit heights are shown in Table~\ref{tab:MCMC} and the fitted models and residuals are displayed in the right columns of Fig.~\ref{fig:mir_models_res}. 

The residuals show significant improvement compared to using the original ALMA vertical height, particularly at shorter wavelengths, however there still exist structures unexplained by the models. This could indicate somewhat different forms in the vertical profile in the mid-infrared compared to the millimeter. Nonetheless, our measurements are able to compare the scale height across these wavelengths under the assumption that the vertical profiles follow broadly the same shape, effectively using the ALMA vertical structure as a ``ruler'' to probe the mid-infrared height. 

To validate the small uncertainties returned by the MCMC samplers, we simulated 100 copies of models at each mid-infrared wavelength with the best-fitting height, adding random noise independently to each pixel in each copy of the model image equal in magnitude to the original observations, before fitting the height characterised by $\phi_\text{B7}$ to each simulated image. At each wavelength, the mean among the 100 best-fit heights were consistent with the value used in the simulations, and the standard deviation among them was comparable to or smaller than the uncertainties on $\phi_\text{B7}$ quoted in Table~\ref{tab:MCMC}. This suggests that while the MCMC uncertainties appear to be very small, they reasonably represent the uncertainties under the assumption that the warped two-component constant $H$ model is reflective of the disk structure. In reality, however, such an assumption may not be fully accurate as reflected by the residuals in Fig.~\ref{fig:mir_models_res}, implying that more realistic uncertainties could be larger, which we discuss further in the next section.

\subsection{Scale height across wavelength}
The weighted mean scale heights from ALMA derived in Section~\ref{sec:almaheight} and the heights from the mid-infrared derived in Section~\ref{sec:mirheight} allow for a consistent comparison of the scale height variations across wavelength. These representative scale heights are listed in Table~\ref{tab:MCMC} and plotted in Fig.~\ref{fig:height}. 

Remarkably, we find that the best-fit heights in the mid-infrared are on average 50\% larger than in the millimeter. It is important to include a discussion on the uncertainties of these measurements. Two sets of uncertainties are shown in Fig.~\ref{fig:height}. The smaller error bars correspond to those na\"ively returned by MCMC fitting and are derived from the uncertainties on $(H_1, H_2, f_1)$ and $\phi_\text{B7}$ listed in Table~\ref{tab:MCMC}. As discussed in Section~\ref{sec:mirheight}, these do not reflect systematic uncertainties in the observations and inaccuracies in the assumed vertical structure to begin with, and is a clear underestimate of more realistic uncertainties given that the large scatter between the mid-infrared wavelengths is unlikely to be real. 

To quantify to first-order a monotonic relationship between scale height and wavelength, we assumed that the two quantities are related by a power law, described by
\begin{equation}
    \label{eq:powerlaw}
    \bar{H}(\lambda) = \left( \frac{\lambda}{\text{1\,mm}} \right) ^p \bar{H}(\text{1\,mm}).
\end{equation}
We fitted $p$ and $\bar{H}(\text{1\,mm})$ to the representative scale height measurements with an MCMC while simultaneously re-estimating the uncertainties on the measurements, which is included as an additional free parameter. The uncertainties on all $\bar{H}$ measurements are assumed to be identical in log space, and represent the most likely uncertainties if Eq.~\ref{eq:powerlaw} were to be true, which could originate from sources such as systematic uncertainties in the observations and inaccurate model assumptions made when measuring the scale heights. 

Such a fit finds that $p = -(8.5 \pm 3.7) \times 10^{-2}$ and $\bar{H}(\text{1\,mm}) = 9.3 ^{+1.5} _{-1.1}$\,au, with the median model and the 1$\sigma$ envelope of models sampled from the posterior distribution shown in Fig.~\ref{fig:height}. The resulting best-fit uncertainties on the $\bar{H}$ measurements are plotted as the larger error bars in Fig.~\ref{fig:height}. These re-estimated uncertainties are significantly larger than those returned directly from MCMC fitting to the images. Even if the relationship defined by Eq.~\ref{eq:powerlaw} is inaccurate, these larger uncertainties still approximate those that would have been derived under the looser assumption that the mid-infrared scale heights should all be similar. We therefore consider these larger and more conservative uncertainties to better reflect the constraints that we are able to place on the scale height from the data analysed in this study.

\section{Discussion}
\label{sec:discussion}

\subsection{Millimeter warp and clumps}
\label{sec:discussionasymmetries}

Prior to this work, the vertical warp in $\beta$~Pic had been detected only at optical/near-infrared wavelengths. Previous work was able to explain the warp with gravitational perturbations on the disk exerted by an inclined planet \citep{Mouillet1997, Nesvold2015}. Such an interaction is thought to take place via secular perturbation, in which the long-term gravitational effect of the planet causes the inclination and longitude of ascending node of disk material to precess at different rates depending on their semimajor axis. The differential inclination precession generates a vertical displacement wave propagating towards the outer edge of the disk on secular timescales, which would appear to vertically puff up the disk. Over time, the wave caused by the different precession rates smears out, leading to a more random distribution within a vertically puffed up disk. 

Following the prediction of such a perturbing planet, $\beta$~Pic~b was discovered via direct imaging \citep{Lagrange2010}, which satisfies the properties required to set off the warp \citep{Dawson2011}. This was considered to be a success in using disk properties to predict planets. 
An additional planet $\beta$~Pic~c had also since then been discovered \citep{Lagrange2019}. Further dynamical simulations \citep{Dong2020} showed that including $\beta$~Pic~c in the picture does not significantly affect the vertical evolution of the disk under secular perturbation, although it may have an effect on eccentricity precession \citep{Smallwood2023}. 

In such a picture, secular perturbation is expected to affect bodies of all sizes, not just the smallest grains probed by scattered light observations. A millimeter warp, for which we see evidence in the ALMA observations, is therefore expected under the secular perturbation interpretation. The consistent location of the warp between scattered light and potentially ALMA observations suggests that they are likely probing the same feature in the disk. 

The dust and CO clump have also been a feature of longstanding investigation. Many potential origins have been proposed, including a clump of debris released from a giant impact \citep{Telesco2005, Dent2014, Jackson2014} or tidal disruption event \citep{Cataldi2018}, planetesimals resonantly trapped by an unseen planet \citep{Wyatt2003, Telesco2005}, secular perturbation due to $\beta$~Pic~b \citep{Nesvold2015} or a collisional avalanche \citep{Grigorieva2007}. 
In our modelling assuming a consistent composition and temperature profile throughout the disk, the minimum grain size required in the dust clump is indeed significantly smaller than in the rest of the disk. Such a difference had been noted in \citet{Telesco2005} relying only on the mid-infrared observations, which our analysis incorporating the constraints in the millimeter confirm. 

Confirmation of such a finding implies that an adequate theory on the origin of the clump should predict an overabundance of small grains in the size distribution or a significantly different composition of the grains. A higher local temperature in the clump to explain its SED is also possible, but is not supported by analyses of the CO temperature \citet{Matra2017}. Catastrophic destruction of planetesimals or planet-sized bodies may be able to disrupt the grain size distribution in its vicinity. 
However, this property alone may not necessarily rule out a resonant planetesimal clump which may also alter the local grain size distribution \citep{Wyatt2008}. Further modelling is required to test whether either class of scenarios could explain the different grain properties between the dust clump and the underlying disk. 

\subsection{Dependence of scale height on grain size}
\label{sec:h(lambda)}

The vertical structure modelling in Section~\ref{sec:vertical} suggests that the disk is vertically thicker in the mid-infrared compared to at millimeter wavelengths. Such a difference is robust, as the clear misfit of the ALMA vertical height at mid-infrared wavelengths (Fig.~\ref{fig:mir_models_res}) indicates that regardless of the specific value fitted for the mid-infrared scale height, the mid-infrared dust distribution is required to be more vertically extended than in the millimeter. Such an effect is unlikely explained by differences in the radial distribution at different wavelengths if the inclination is slightly non-edge-on, as its effect would be the opposite: while we would see the radial distribution contribute to the apparent vertical distribution in that case, the mid-infrared emission probes radially closer-in dust, whereas we find the disk to be even more vertically extended in the mid-infrared.

As thermal emission from debris disks at a given wavelength primarily originates from grains at a similar size to the wavelength, our finding suggests that the distribution of micron-sized grains is vertically puffier than mm-sized grains, which in turn suggests a larger inclination dispersion in micron-sized grains. 

What could be the reason for the significant difference in scale height between the two grain size populations? None of the effects mentioned in Section~\ref{sec:introduction} commonly studied are expected to directly puff up smaller grains significantly more than larger grains. 
Collisional damping dissipates the kinetic energy of successively smaller grains produced from larger ones, reducing mutual particle inclinations. The damping rate is dominated by collisions with similarly sized grains. In debris disks, smaller bodies are expected to have a larger collective cross-sectional area than larger ones, which makes their collisional damping rate higher. As a result, if collisional damping is an important effect, smaller dust grains should be more settled around the disk midplane \citep{Pan2012}, which is contrary to what we observe. Recent kinetic modelling that incorporates fragmentation has also suggested that collisional damping is inefficient for millimeter-sized grains under common material strength assumptions \citep{Jankovic2024}. 
Viscous stirring and dynamical friction may pump up the inclination dispersion of smaller bodies while damping larger stirring bodies, however the dynamical heating of dust grains due to dynamical friction is expected to be negligible \citep{Goldreich2004, Pan2012}. 

One potential source for the negative correlation between inclination dispersion and grain size may relate to radiative effects. Stellar radiation pressure is thought to set the minimum grain size in the disk, below which dust grains are put on hyperbolic orbits and removed from the system. Just above this blowout limit, stellar radiation pressure increases the eccentricity of small grains, which may be converted into an inclination dispersion through random collisions between these small grains \citep{Thebault2009}. As smaller grains have larger $\beta = F_\text{rad}/F_\text{grav}$ and are thus more affected by radiation pressure, a grain size dependence of the inclination dispersion consistent with the trend which we find is expected. 

To estimate the expected scale height induced by radiation pressure, we first estimated the characteristic grain size that contribute to the observed emission at each wavelength. Although this characteristic grain size should be close to the observing wavelength, in practice it could be a factor of a few smaller, particularly at the mid-infrared wavelengths compared to millimeter wavelengths. Using the same dust emission model (including parameters defining the composition and size distribution, with $D_\text{min} = 1 \mu$m) as in Section~\ref{sec:sed}, we find that at 100\,au where the radial profile peaks in the millimeter, 
the central 1$\sigma$ grain size range contributing to emission is 1.2--2.2\,$\mu$m at $\lambda = 8.8 \, \mu$m, 1.4--6.3\,$\mu$m at $\lambda = 24.5 \, \mu$m, 0.8--46\,mm at $\lambda = 0.87$\,mm and 1.4--87\,mm at $\lambda = 1.3$\,mm. This implies that the mid-infrared emission is dominated by dust close to the minimum grain size over a factor of 5 in size, with $\beta$ ranging between 0.4 and 3, while the mm emission is dominated by much larger dust over a factor ~50 in size (and $\beta$), where $\beta<0.003$).
Under the \citet{Thebault2009} model, the inclination dispersion induced by radiation pressure reaches a steady state of approximately 8$^\circ$ for $\beta$ between 0.3 and 0.4, which corresponds to a scale height of 14\,au at a radius of 100\,au, consistent with the mid-infrared height that we measure. At millimeter wavelengths, the inclination dispersion due to radiation pressure is small ($\ll$1$^\circ$) and is not expected to be the main contributor to the vertical height.)

Given the presence of gas detected in $\beta$~Pic (including CO, \citealp{Dent2014, Matra2017}, C\textsc{i}, \citealp{Cataldi2018} and O\textsc{i}, \citealp{Brandeker2016}), it is possible that gas could play a role in shaping dust dynamics if molecular hydrogen as a remnant from the protoplanetary disk phase is also present in the disk at a sufficiently high level. 
\citet{Olofsson2022} found that as gas causes grains to settle to the midplane, decreasing the height of debris disks, the effect is in fact stronger for smaller grains such that micron-sized dust should have a smaller scale height than millimeter-sized dust. However, if the gas were to be turbulent, as discussed in \citet{Marino2022}, smaller, micron-sized grains could be vertically stirred by turbulence. In this scenario, both micron- and millimeter-sized dust grains could be at equilibrium between turbulent stirring and vertical settling, at which smaller grains have larger scale-heights \citep{Youdin2007}. This is a common assumption in gas-rich protoplanetary disks \citep{Armitage2020}. 

To evaluate the plausibility of this scenario, we estimated whether the CO mass detected could feasibly produce the vertical height variation across wavelength. The relation between the dust and gas vertical height can be approximated as 
\begin{equation}
    \label{eq:h_dust_gas}
    \frac{H_\text{dust}(a)}{H_\text{gas}} \approx \sqrt{\frac{\alpha}{\alpha + \mathrm{St}(a)}} \ ,
\end{equation}
where $a$ is the diameter of a dust grain, $\alpha$ is the Shakura-Sunyaev viscosity parameter and $\mathrm{St} = \Omega_K t_s$ is the Stokes number, where $\Omega_K$ is the Keplerian angular speed and $t_s$ is the stopping time of the dust grain due to gas drag \citep{Youdin2007}. 

The low CO mass in the disk (which is estimated to be $3.4^{+0.5}_{-0.4} \times 10^{-5}$M$_\oplus$, \citealp{Matra2017}) implies a significantly larger mean free path than the dust sizes considered here, such that the gas drag stopping time can be estimated in the Epstein regime with
\begin{equation}
    t_s = \frac{\rho_\mathrm{dust}}{\rho_\mathrm{gas}} \frac{a}{c_s},
\end{equation}
where $\rho_\mathrm{dust}$ is the dust grain density (assumed here to be 2.7\,g/cm$^3$), $\rho_\mathrm{gas}$ is the gas (mass) density and $c_s$ is the speed of sound. If the gas is secondary in origin, consisting primarily of collisionally released CO with a mean molecular weight of 28, the gas density is so low that $\mathrm{St}$ becomes too large for effective interaction with even micron-sized dust in the outer disk ($\mathrm{St}\sim10^5$). 

If we were to instead assume that the gas is primordial, i.e., consisting primarily of H$_2$ with a mean molecular weight of 2.4, has a total gas mass $4 \times 10^5$ times larger than the CO mass (based on the lower end of the CO/H$_2$ abundances measured in protoplanetary disks, \citealp{Trapman2025}) and is uniformly distributed between 20 and 100\,au (although in reality the CO distribution is observed to be asymmetric), then we find $\mathrm{St}$ to be $\sim0.06$ for 5\,$\mu$m grains and $\sim 10$ for 800\,$\mu$m grains, which are the grain sizes probed by the shortest and longest wavelength observations in this study. Eq.~\ref{eq:h_dust_gas} then implies that the $\alpha$ value required to explain the $H_\mathrm{dust}(\mu\mathrm{m})/H_\mathrm{dust}(\mathrm{mm}) \sim 1.5$ observed is $\sim 8$, which is unrealistically large compared to typical values of $10^{-2}$ to $10^{-5}$ measured in protoplanetary disks \citep{Longarini2025} and would imply supersonic turbulence. 

To explain the dust height variation with a more realistic $\alpha$ of $\sim 10^{-2}$ would require an H$_2$ mass that is $\sim 10^8$ times larger than the observed CO mass, implying a gas disk mass of $\sim 0.01 \, M_\odot$. 
This is not an unrealistically large mass if $\beta$~Pic is harboring remnants of its protoplanetary disk, as it is an order of magnitude smaller than recently determined dynamical masses of a sample of protoplanetary disks \citep{Longarini2025}. 
While this scenario cannot be fully ruled out considering the young age of $\beta$~Pic and the trend of decreasing CO abundance from the ISM to protoplanetary disks (albeit on the more extreme end of plausible ranges), it contradicts observational upper limits on the H$_2$ mass of $3 \times 10^{-4} \, M_\mathrm{Jup}$ placed by far-ultraviolet observations \citep{LecavelierdesEtangs2001}. In other debris disks with CO detections, H$_2$ has also been constrained to be no more than $\sim 10^3$ times more abundant than CO in HD\,110058 and $\sim 10^4$ times in HD\,131488 \citep{Smith2025}. 

Furthermore, if gas were to be dominant in setting the vertical height of dust, the gas height would be at least as large as the dust height. \citet{Matra2017} found the CO scale height to be 7\,au at a radius of 85\,au, which is the de-projected radial location of the CO clump based on ALMA velocity information. This is smaller than the apparent dust height observed both by ALMA and in the mid-infrared, requiring a scenario in which the gas boosts dust grains to heights beyond its own reach, which is unlikely. For comparison, the theoretical scale height of CO assuming the blackbody equilibrium temperature at 85\,au is $c_s / \Omega_K = 3\,\mathrm{au}$. However, it should be noted that the parametrisation for the CO vertical profile is different from what we assumed for dust in this study, so the values may not be readily comparable. The measured CO scale height may also be smaller than the true vertical reach of all gas species \citep{Marino2022}, as the upper layers of CO could have photodissociated into atomic C (which has not been vertically resolved, \citealp{Cataldi2018}) and O. The gas explanation for vertically puffier small grains would therefore simultaneously require an unusually large H$_2$ reservoir in comparison to CO and a vertically stratified distribution of species, with the overall gas disk being vertically thicker than CO alone. 

Finally, a grain size dependence of the scale height like that inferred might also be a result of a larger proportion of small grains being found within the tilted secondary disk relative to the primary disk. Model residuals in such a scenario should be vertically asymmetric and reflect the tilt of the secondary disk. 
While vertical asymmetries could be present at the longer wavelengths in Fig.~\ref{fig:mir_models_res}, there is no significant evidence of this effect at the shorter wavelengths, suggesting that an overabundance of small grains in a non-precessing secondary disk is unlikely to explain all of the larger scale height measurements in the mid-infrared. 

While \citet{Vizgan2022} found the aspect ratio of AU Mic to be larger in Band 6 compared to Band 9, we found no significant difference between the scale height of $\beta$~Pic between Bands 6 and 7 studied here, and the larger mid-infrared height that we find is opposite to the wavelength dependence suggested for AU Mic at millimeter wavelengths. It is possible that the scale height exhibits a non-monotonic dependence on the wavelength, which will require further observations to explore. 
Future work may also wish to consider the scale height in scattered light. Previous scattered light modelling has found a large range of best-fit values, including $h = 0.02$ at an inclination of 87.7$^\circ$ \citep{Ahmic2009}, $h = 0.044$ at 86$^\circ$ \citep{Milli2014}, $0.05 < h < 0.1$ at 85--87$^\circ$ \citep{Kalas1995} and $h = 0.100 \pm 0.005$ at $h = 84.6 \pm 0.03^\circ$ \citep{Milli2026}. Consistent multi-wavelength modelling will be required to attempt to draw a robust comparison between the vertical structure seen in thermal emission and scattered light.

\subsection{Constant height across radius}
\label{sec:constantheight}
Having discussed scale height as a function of grain size, let us now turn to the question of scale height as a function of radius. 

Models of debris disk observations have often assumed the vertical height to be proportional to radius, i.e., having a constant ``aspect ratio'', which corresponds to the assumption that the inclination dispersion is constant for bodies at all radii. This is unlike gas-rich protoplanetary disks in which the gas scale height increases faster than radius (i.e., ``flaring'', \citealp{Chiang1997}). In debris disk models by \citet{Kenyon2001}, in which the scale height is predominantly set by viscous stirring, the resulting eccentricity and inclination dispersion indeed remains largely constant across radius. However, the relatively constant height observed in $\beta$~Pic suggests an inclination dispersion that decreases as $r^{-1}$, in contrast to the $h \propto r^{-0.3 \pm 0.2}$ found by \citet{Matra2019} which is more similar to a constant aspect ratio.

It is possible that large quantities of gas, if present, could affect the vertical distribution of dust. Simulations by \citet{Olofsson2022} without turbulence found that gas drag pushes particles towards larger radii, causing them to settle to the midplane as they experience a headwind every time they cross the midplane. 
A sufficiently high gas mass (on the order of 10$^{-2}$M$_\oplus$) could result in a vertical height that decreases with radius further out from the parent ring, where grains have been displaced by radiation pressure and gas drag.
However, as millimeter grains do not experience significant radial displacement due to radiation pressure, it is not clear whether in the absence of radial migration, settling alone could result in a constant height across radius. 

Another explanation could involve dynamical effects due to $\beta$~Pic~b and c. Although our scale height modelling already accounted for the warped midplane, the effect of the planets on the disk could be more complicated. Due to the young age of $\beta$~Pic at 21~Myr, the vertical displacement waves set off by secular perturbations from the planets may have not yet puffed up the entire disk, with its effects limited primarily to the inner and intermediate regions of the disk where the warp is currently seen. Interestingly, the 100\,au transition radius in our model, where the effective position angle of the inner and outer disk changes, is also the semimajor axis where the secular timescale \citep{Murray1999} due to the most massive planet in the system known, $\beta$~Pic~b \citep{Lagrange2020}, is equal to the age of the system. Particles outside this radius could still remain free from these secular effects, thereby exhibiting a lower inclination dispersion in the outer disk. A somewhat constant apparent scale height formed as such has been suggested by secular forcing models \citep{Dawson2011}. Recent models that account for the disk's self-gravity have suggested that both massless and massive debris disks could exhibit a scale height that increases with radius before decreasing: in the massless case this occurs before the warp fully propagates through the disk, whereas in the massive case this occurs due to the disk warp becoming localised \citep{Sefilian2025}. The requirements in either scenario are likely satisfied. The secular reach of $\beta$~Pic~b (the impact of the less massive and closer-in planet c is negligible) is only 100\,au given the age of the system, whereas the disk mass inferred from collisional cascade models ($\sim100\,M_\oplus$, \citealp{Pan2012, Zawadzki2026}) is comparable to disk mass requirements for a localised clump to occur \citep{Sefilian2025}. Either scenario would in theory produce a sharp transition in aspect ratio within the disk, which would be more complex than can be described by a single power-law. It is possible that our constant height model offers an approximation that yields a reasonable fit, and further modelling would be required to investigate this scenario in more detail.

\subsection{Vertical profile shape}
With the inclusion of ALMA Band 7 data in this study, we confirm the finding in \citet{Matra2019} based on Band 6 data that the vertical structure of $\beta$~Pic is non-Gaussian. Recent modelling of high-resolution ALMA observations \citep{Marino2026} have also found a majority of debris disks in their sample to be better described by a Lorentzian vertical profile \citep{Zawadzki2026}, in contrast to a common model assumption that debris disks are vertically Gaussian. We briefly discuss choices for non-Gaussian parametrisations and possible interpretations in this section. 

\citet{Matra2019} showed analytically that for a Rayleigh distribution in inclination, the resulting scale height is Gaussian provided that the inclination dispersion is small. The assumption of a Rayleigh inclination distribution is applicable to self-stirred disks in which bodies within the disk gravitational excite other bodies to provide inclination and eccentricity \citep{Lissauer1993, Stewart2000}. To model the non-Gaussian vertical profile of $\beta$~Pic, \citet{Matra2019} found a double-Gaussian model to offer an appropriate fit, which was motivated by the dynamically hot and cold population in the Solar System's Kuiper belt. While we confirm in this study that a double-Gaussian model is able to reproduce the ALMA observations reasonably well at both bands, other parametrisations are also available, which have been explored in \citep{Matra2019} and other studies. 
In particular, in scattered light imaging, \citet{Olofsson2022} were able to produce relatively close fits to a sample of highly inclined debris disks assuming a vertical distribution defined by a super-Gaussian function,
\begin{equation}
    \xi(z) \propto \exp \left[ - \left( \frac{z^2}{2 H^2} \right)^k \right],
\end{equation}
\noindent in which $H$ is the scale height at a given radius. Setting $k = 1$ recovers a Gaussian function. Such a parametrisation has also been used to specifically model scattered light observations of $\beta$~Pic \citep{Kalas1995}, which found the best fitting value of $k$ to be smaller than 1. 

To compare between these parametrisations, we plotted super-Gaussian functions for a few cases of $k$ in Fig.~\ref{fig:vertical_profiles} for comparison. Compared to our best-fit profiles, a regular Gaussian produces a peak that is too broad and wings that are too narrow. Reducing $k$ sharpens the peak, and an approximation to the best-fit profiles is obtained with $k \approx 0.5$. This is consistent with the best-fit super-Gaussian in \citet{Matra2019}, which found $k\approx 0.45 \pm 0.05$ (converted from the $p$ parameter in their equivalent form of this parametrisation).

One possible origin of the double-Gaussian-like vertical distribution could be a self-stirred planetesimal belt with two dynamical populations, as proposed by \citet{Matra2019}. However, the scenario of incomplete secular perturbation in the disk, as discussed in Section~\ref{sec:constantheight}, would add further complexity to this picture, as the inclination dispersion is expected to be large in the inner regions of the disk and small in the outer regions. The essence of the vertical profiles shown in Fig.~\ref{fig:vertical_profiles} is a Gaussian distribution but with an added pointy peak. The apparent non-Gaussian profile could be reflective of the combined projected effect of a thick inner disk and thin outer disk, where the thin outer disk contributes to the pointy peak and remains to be puffed up significantly from secular perturbation due to the slower precession timescales at large semimajor axes. For disks with nonzero mass, a localised warp produced by the combined effect of an inclined perturbing planet and the disk's self-gravity, as discussed in Section~\ref{sec:constantheight} to explain the radial dependence on the scale height, has also been suggested to produce non-Gaussian vertical profiles, similar to the ones we fitted, in the regime where the ratio between the disk and planet mass is approximately less than unity \citep[see e.g. their Fig. 4]{Sefilian2025}. This mass ratio based on the disk mass \citep{Zawadzki2026} inferred from collisional cascade models \citep{Pan2012} and the mass of planet b \citep{Lagrange2020} is a few percent. 

It has also been suggested that a non-Gaussian vertical profile could arise from the combined emission from a range of grain sizes, each normally distributed, but with a different scale height dependent on the grain size \citep{Terrill2023}. 
The wavelength dependence of scale height that we find is shallow (Eq.~\ref{eq:powerlaw}), which would result in a vertical profile well-approximated by a Gaussian. The grain size effect alone thus appears unlikely to offer a full explanation of the vertical non-Gaussianity observed in $\beta$~Pic. 

Ongoing work is also investigating Lorentzian vertical profiles created by out-of-equipartition stirring, in which small bodies in the disk, starting with a significantly higher eccentricity dispersion relatively to the inclination dispersion, enter into an era of lognormally distributed inclinations due to encounters with large bodies in the midplane, which translates to Lorentzian vertical density profiles, before eventually establishing equipartition, characterised by Rayleigh-distributed inclinations and, correspondingly, Gaussian vertical density distributions \citep{Chiang2026}. 

\subsection{Presence of small grains}
Our SED modelling in Section~\ref{sec:sed} suggests the presence of grains as small as 1~$\mu$m within the disk, which is 5 times smaller than the radiation pressure blowout limit. To explain such an abundance of small grains, there either exist mechanisms that inhibit or slow down their removal from the system, or we are witnessing the disk at a special time soon after the production of a large population of small grains.

Gas in the disk could slow down the removal of small grains produced via a steady-state collisional cascade, however it is not clear whether the gas density \citep{Dent2014, Matra2017}, with large uncertainties as discussed in Section~\ref{sec:h(lambda)}, would be sufficient to allow for the level of buildup to be reflected in the SED. 

The recent discovery in the mid-infrared of the tilted secondary disk to the NE and the extended filamentary dust structure, or cat's tail, to the SW with JWST/MIRI has lead to the proposal of a model in which both features may be produced by small dust released from the destruction of a planetesimal, some of which are on their way to being removed from the system \citet{Rebollido2024}. Such a scenario relates closely to potential origins of the SW dust clump discussed in Section~\ref{sec:discussionasymmetries}, within which the smallest grains required to be at 0.2~$\mu$m in our SED models are even smaller than in the rest of the disk. 

\citet{Rebollido2024} proposed that sporadic production of sub-blowout sized dust could appear to spiral away from the system on a curved trajectory to form the cat's tail when projected, which in their model is $\sim$150~yr old. However, striated features are present across the MIRI image of $\beta$~Pic, which the authors speculate could be remnants of similar events in the past. Significant spectral variations observed in $\beta$~Pic over the timescale of two decades has added further evidence for planetesimal collisions in the inner system followed by the blowout of small grains \citep{Chen2024}. If such events producing small dust are indeed common, it would be more likely for significant quantities of sub-blowout sized dust to be present in the system. 

It is possible for the grain composition assumed and imperfections of grain emission models to affect the fitted minimum grain size. Here we assumed the dust composition to be free of ice, finding a minimum grain size below the radiation pressure blowout size. \citet{Marshall2026} similarly assumed an ice-free composition when fitting to the SED of a sample of debris disks and found minimum grain sizes in A stars to be commonly smaller than the blowout size. These results could depend on the water ice fraction assumed \citep{Sommer2025}, and previous work have suggested that assuming a larger water content could lead to a larger minimum grain size inferred among Sun-like stars, or a smaller minimum grain size for more luminous stars \citep{Stuber2022}. The size distribution may also depart form a perfect power law, and could instead exhibit wavy patterns which we have not modelled here \citep{Thebault2007}.

\subsection{Dynamical scenario in $\beta$~Pic}
\label{sec:scenario}

The many features in $\beta$~Pic have long presented a challenge to explanation with a consistent dynamical scenario. With the ever more abundant observations and theoretical progress on this archetypal system, we may be approaching a stage where we could renew our attempts to put together such a picture. 

Without detailed dynamical simulations linking all features, it is difficult to lay out exactly what such a scenario could look like. However, the picture emerging from these observations appears to be one in which dust production cannot be fully explained with only a generic steady-state collisional cascade. Given the prominent role of $\beta$~Pic~b and c in the system, the process could have initially been set off by the relatively well-studied effects of secular perturbation from the two giant planets interior to the disk \citep{Mouillet1997, Nesvold2015, Smallwood2023}, manifesting as a warp seen in both scattered light and the millimeter while puffing up the inner and intermediate regions of the disk more than the outer regions. 

The vertical structure of the disk is then no longer set by the self-stirring of bodies within the disk, leading to a non-Rayleigh distribution in inclination \citep{Matra2019}. Instead, a non-Gaussian vertical distribution is introduced, potentially due to the complex disk geometry caused by secular forcing, including the as-yet relatively unaffected outer regions of the disk contributing to enhanced projected emission concentrated near the midplane, which future studies may wish to investigate. 

In addition to effects on the vertical structure, such a secular wave could have also enhanced collision rates between planetesimals in the disk, triggering collisions between large bodies on dynamically excited, inclined orbits. This is consistent with JWST/MIRI MRS observations that point recent giant collisional activity \citep{Chen2024}. In the more distant past, such collisions could have involved a giant impact catastrophically destroying a Mars-sized body on an inclined orbit within the past 0.5~Myr \citep{Jackson2014}, releasing material onto a range of orbits which all pass through the initial point of collision. This would have resulted in a stationary dust clump at the original collision point on the SW side of the disk as we observe in the mid-infrared and now possibly in the millimeter, and potentially an extended secondary disk overlaid on the warp on the NE side as we observe with MIRI \citep{Jones2023}, where the orbits of the debris smear out over a large range as they reach apastron. The details of how enhancements in collision rates could be achieved with secular perturbation and detailed modelling of the outcome of a giant impact in $\beta$~Pic could be subjects for future studies. 

Among this tilted quasi-steady-state disk of debris, secondary collisions between large fragments occur sporadically to release transient dust tails \citep{Rebollido2024}, one of which could have occurred recently enough in the past 150~yr as proposed to explain the cat's tail seen by MIRI. However, small dust could be plentiful in a disk with such a high collision rate. While the gas mass may not be high enough to dictate their dynamics \citep{Olofsson2022}, radiation pressure excites their eccentricity, which is converted to an inclination dispersion via collisions \citep{Thebault2009}, causing small dust to be puffed up vertically, possibly also aided by gas effects if the molecular hydrogen were to be abundant and the gas were to be turbulent \citep{Marino2022}. Future work may wish to investigate whether the scale height and grain size trend observed in this study could be achieved via these mechanisms. 

Gaseous species could also have been released from the giant impact that produced the SW clump and NE secondary disk, contributing to the observed asymmetric and tilted CO emission aligned with the clump, both of which appear to be vertically offset from the midplane \citep{Matra2017, Skaf2023} in a direction consistent with the tilt of the secondary disk. If the body being disrupted was planet-sized, a large amount of the CO could have come from its atmosphere, releasing a CO mass comparable to the $10^{-5}$~M$_\oplus$ of CO found around HD~172555 which is thought to be stripped from a giant impact of a planet in the system \citep{Schneiderman2021}. The dust clump, which contains the greatest abundance of small grains through enhanced collision rates at the point where debris orbits meet, releases sub-blowout sized dust which could be carried by gas from the CO clump to further enrich the small grain population throughout the disk. Radiation pressure will still eventually remove the sub-blowout sized dust from the system, with the interaction between dust and gas possibly smearing out the CO distribution to appear broader at the clump than an unresolved collision point.  

Such a highly speculative scenario clearly involves several key steps that require future work to investigate as outlined above. Nonetheless, this may point out a few key areas of study that future work may wish to investigate. 
Firstly, dynamical simulations are required to test whether the rate of collisions or close encounters enabling tidal disruption is indeed increased due to secular waves in the disk. Secondly, such simulations may also be able to show whether the non-Gaussian shape of the vertical profile observed as well as its apparent lack of dependence on radius can indeed be reproduced by secular perturbation. Thirdly, detailed simulations are required to study whether radiation pressure and random collisions can result in the grain size dependence of scale height observed and whether turbulent gas could contribute to this dependence. There has recently been observational evidence in non-debris disk systems of radiation pressure accelerating micron-sized dust away from the star while gas is carried along with this dust \citep{Han2022b}, and simulations have suggested that radially migrating dust grains in debris disks could radially broaden the gas pressure maximum \citep{Weber2026}, but this remains to be tested for the physical conditions in $\beta$~Pic.

\subsection{Future work on the vertical height}
To understand the generality of the significant difference in scale height between micron and mm-sized grains, it is worth pursuing further observations of the scale height as a function of wavelength for a wider range of disks. While it is generally difficult to vertically resolve debris disks in the mid-infrared, this may be achievable for a small sample of systems with JWST/MIRI, and inclined disks with currently available JWST observations (e.g., Fomalhaut, \citealt{Gaspar2023} and $\gamma$~Oph, \citealt{Han2026}) could be studied with consistent multi-wavelength modelling. Upcoming MIRI non-coronagraphic imaging of $\beta$~Pic will also be well-suited for carrying out this analysis \citep{Perrin2024}. 
 
Several additional avenues may also be taken. Firstly, more comprehensive data analysis methods may be developed to be consistently applied across scattered light data in which a much larger number of disks may be detected and vertically resolved. In particular, this may require methods to better account for the scattering phase function, which may deviate from common parametrisations assumed such as the Henyey-Greenstein scattering phase function. 

Secondly, using different ALMA bands separated as widely as possible may be able to probe any vertical height variations within the millimeter wavelength regime. This study did not find any significant differences in scale height between the Band 6 and 7 images, but \citet{Vizgan2022} suggested the vertical height of AU~Mic to be different between Bands 6 and 9. As their best-fit scale heights just overlap within the 1$\sigma$ uncertainty range reported, extending this analysis to bands at wider wavelength separations could test the generality of this finding. If resolved in Bands 1 and 9, for example, the factor of 16 difference in wavelength would correspond to a factor of 1.27\,$\pm$\,0.13 in scale height according to Eq.~\ref{eq:powerlaw}. A comparative analysis between debris disks with and without gas could also yield insight into the impact of gas on the vertical dust height \citep{Hales2022}. 

The results in this study also suggest that effects that differentially puff up the disk for different dust populations may be worth taking into account for dynamical models that aim to account for the vertical disk structure, possibly including effects derived from stellar radiation pressure. 
In light of the wealth of observations of $\beta$~Pic, simulations may also wish to explore consistent dynamical scenarios causing the multiple features now found to exist in the disk, such as the type outlined in Section~\ref{sec:scenario} or alternative possibilities.

\section{Conclusions}
\label{sec:conclusions}
The main conclusions of this study are summarised as follows. 

\begin{enumerate}

\item $\beta$~Pic is currently the only debris disk in which the scale height has been compared across a range of very widely separated wavelengths in thermal emission that span a factor of $\sim$150. We find the mid-infrared scale height to be on average 1.5 times the millimeter height. This could occur via the combined effect of radiation pressure and random collisions, which acts to vertically puff up smaller grains more than larger grains. Turbulent gas could also contribute to this trend, although the low CO gas mass detected and the H$_2$ upper limit make this scenario less likely. The scale height is not significantly different between ALMA Bands 6 and 7.

\item We found the disk to be vertically warped in the millimeter. There is tentative but insignificant evidence of the 50\,au SW dust clump and possibly 30\,au NE secondary dust clump in the millimeter which were previously imaged in the mid-infrared. Compared to the SW side, the NE side of the disk also appears brighter at 100 to 200\,au, where a tilted secondary disk has been detected in the mid-infrared. 
The presence of these features at both mid-infrared and millimeter wavelengths is expected if the warp were to be produced via secular perturbation from the two giant planets in the system, and radial asymmetries such as clumps via giant impact debris. 

\item Our modelling suggests that the disk is more adequately fitted by a model in which the height is constant across radius, as opposed to being proportional to radius as is commonly assumed in debris disks. A vertical profile parametrised by the sum of two Gaussians can largely reproduce the observations once the constant height and warp are accounted for in our models. 

\item Our mid-infrared to millimeter SED models of spatially resolved regions in the disk require the presence of micron-sized grains smaller than the radiation pressure blowout limit in the disk, and a population of dust grains in the dust clump that is even smaller in size or of different composition. 

\item In addition to the vertical warping, a secularly perturbed disk might also explain the more constant $H(r)$ and non-Gaussian vertical profile in the disk, which detailed modelling in future work may test. We hypothesise that a giant impact could be responsible for the production of the primary dust clump and secondary disk, within which subsequent collisions could release small grains into the disk. Future studies may also wish to investigate whether such a giant impact could have been linked to changes in collision rates caused by secular precession waves in the first place. 

\end{enumerate}

\begin{acknowledgements}
We are grateful to Chat Hull for providing the calibrated visibilities of the Band 7 observations used in this study. 
Y.H. is funded by a Caltech GPS Barr Fellowship. 
This study is supported in part by a Gates Cambridge Scholarship from the Gates Cambridge Trust enabled by Grant No. OPP1144 from the Bill and Melinda Gates Foundation. 
M.R.J. acknowledges funding provided by the Institute of Physics Belgrade through the grant by the Ministry of Science, Technological Development, and Innovations of the Republic of Serbia.
A.M.H. is supported by the National Science Foundation under Grant No. ASTR-2307920. 
L.M. acknowledges funding by the European Union through the E-BEANS ERC project (grant No. 100117693). Views and opinions expressed are however those of the author(s) only and do not necessarily reflect those of the European Union or the European Research Council Executive Agency. Neither the European Union nor the granting authority can be held responsible for them.
This paper makes use of the following ALMA data: ADS/JAO.ALMA\#2012.1.00142.S, 2019.1.00041.S. ALMA is a partnership of ESO (representing its member states), NSF (USA) and NINS (Japan), together with NRC (Canada), NSTC and ASIAA (Taiwan), and KASI (Republic of Korea), in cooperation with the Republic of Chile. The Joint ALMA Observatory is operated by ESO, AUI/NRAO and NAOJ. The National Radio Astronomy Observatory and Green Bank Observatory are facilities of the U.S. National Science Foundation operated under cooperative agreement by Associated Universities, Inc.
This research made use of NASA's Astrophysics Data System; the \textsc{IPython} package \citep{ipython}; \textsc{SciPy} \citep{scipy}; \textsc{NumPy} \citep{numpy}; \textsc{matplotlib} \citep{matplotlib}; and \textsc{Astropy}, a community-developed core Python package for Astronomy \citep{astropy}. 
The T-ReCS data used in this study are available on the Gemini Observatory Archive under programme IDs GS-2003B-Q-14 and GS-2010B-Q-50. The VISIR data used in this study are available on the ESO Archive under programme ID 095.C-0425(A). The ALMA data used in this study are listed on the ALMA Science Archive under programme ID 2012.1.00142.S for the Band 6 observations and 2019.1.00041.S for Band 7. 
\end{acknowledgements}

\bibliographystyle{aa}
\bibliography{references.bib}

\begin{appendix}
\section{Mid-infrared asymmetries}

\label{appendix}

\begin{figure*}
    \centering
    \includegraphics[width=17cm]{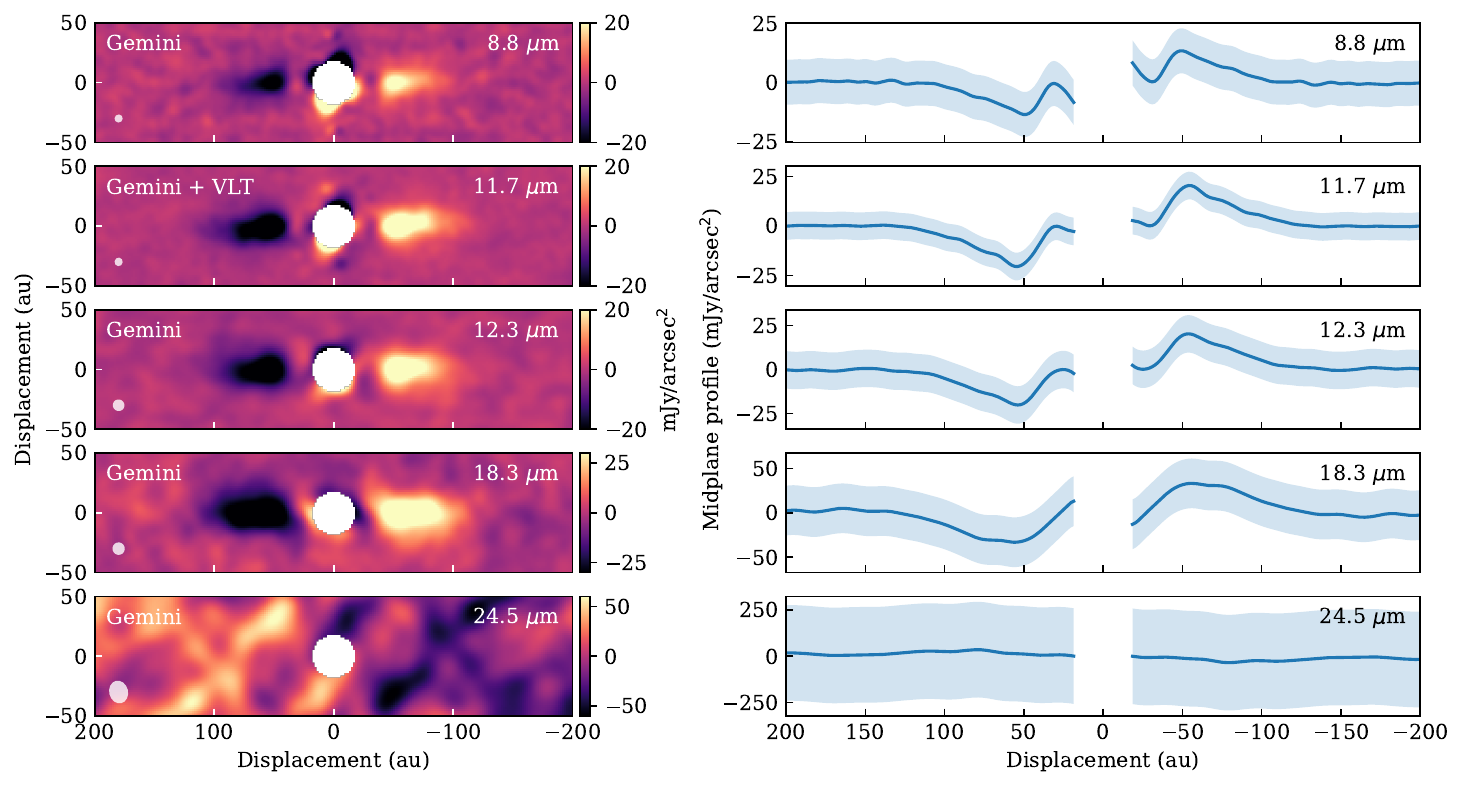}
    \caption{Left panels: rotationally subtracted mid-infrared images of $\beta$~Pic obtained by subtracting each image rotated by 180 about the star from the original image to emphasize any asymmetric features in the disk. The inner regions of the mid-infrared observations are masked out as their self-subtractions are dominated by PSF asymmetries. The images are smoothed with Gaussian kernels with FWHMs of 8.3, 8.3, 12.5, 12.5 and 20.8\,au from the shortest to longest wavelength respectively to achieve the effective PSFs indicated by the ellipses. The smoothed images are displayed on linear scales floored/capped between $\pm$0.02, 0.02, 0.02, 0.03 and 0.6~mJy/arcsec$^2$ from the shortest to longest wavelength respectively. 
    Right panels: Mean projected surface brightness profile of the smoothed and rotationally subtracted images within 20\,au from the disk's major axis. The uncertainties are estimated from an analogous profile of an empty region in the background of each image.}
    \label{fig:image180_mir}
\end{figure*}

This appendix displays the rotationally subtracted mid-infrared images in Fig.~\ref{fig:image180_mir}. The images were centred by fitting a 2D Gaussian model to the stellar PSF. 

\end{appendix}
\end{document}